\documentclass[twoside]{article}

\usepackage{times}

\usepackage[latin1]{inputenc}
\usepackage{a4}
\usepackage{tabularx}
\usepackage{epsf}

  \usepackage{graphicx}
\DeclareGraphicsExtensions{.gif,.eps,.jpg,.png}

\def\bref{\vspace{4pt}\noindent\hangindent=10mm}

\graphicspath{{plots/}}

\begin{document}

\def\myfigure#1{\noindent\parbox{\textwidth}{%
\rule{\textwidth}{0mm}\\[1mm]\centerline{#1}}\\[4mm]}

\newdimen\mywidth
\newcounter{mycount}
\def\arabiclist{
\settowidth\mywidth{--(0)--}
\begin{list}
  {(\arabic{mycount})}
  {\usecounter{mycount}
   \leftmargin\mywidth
   \labelwidth\mywidth
   \topsep0ex}}

\def\HII{H\,{\sc ii}}
\def\HeI{He\,{\sc i}}
\def\CIV{C\,{\sc iv}}
\def\NIV{N\,{\sc iv}}
\def\NV{N\,{\sc v}}
\def\OI{O\,{\sc i}}
\def\NeIII{Ne\,{\sc iii}}
\def\NeII{Ne\,{\sc ii}}
\def\SiII{Si\,{\sc ii}}
\def\SiIV{Si\,{\sc iv}}
\def\SII{S\,{\sc ii}}
\def\ArIII{Ar\,{\sc iii}}
\def\ArII{Ar\,{\sc ii}}
\def\CaII{Ca\,{\sc ii}}
\def\MgII{Mg\,{\sc ii}}
\def\FeIII{Fe\,{\sc iii}}
\def\FeV{Fe\,{\sc v}}
\def\Teff{T_{\rm ef\kern-.05em f}}
\def\logg{\log g}
\def\vinf{v_\infty}
\def\Mdot{\dot M}
\def\Msun{M_\odot}
\def\Zsun{Z_\odot}
\def\Rsun{R_\odot}
\def\Lsun{L_\odot}

\pagestyle{myheadings}
\markboth{\hspace*{\fill}{\it \small
A.\,W.\,A. Pauldrach
}}
{{\it \small
Hot Stars: Old-Fashioned or Trendy?
}\hspace{\fill}}

\vspace*{-1.4cm}
\thispagestyle{empty}

\vspace*{1.7cm}

\begin{center}
{\Large\bf Hot Stars:\\[0.2cm]
           Old-Fashioned or Trendy?}\\[0.7cm]

A.\,W.\,A.~Pauldrach \\[0.17cm]
Institut f\"{u}r Astronomie und Astrophysik der
Universit\"{a}t M\"{u}nchen\\
Scheinerstra{\ss}e~1, 81679~M\"{u}nchen, Germany \\
{\tt UH10107@usm.uni-muenchen.de},
{\tt http://www.usm.uni-muenchen.de/people/adi/adi.html}
\end{center}

\vspace{0.2cm}

\begin{abstract}
\noindent{\it
Spectroscopic analyses with the intention of the interpretation of
the UV-spectra of the brightest stars as individuals -- supernovae
-- or as components of star-forming regions -- massive O~stars --
provide a powerful tool with great astrophysical potential for the
determination of extragalactic distances and of the chemical
composition of star-forming galaxies even at high redshifts.

The perspectives of already initiated work with the new generation
of tools for quantitative UV-spectroscopy of Hot Stars that have
been developed during the last two decades are presented and the
status of the continuing effort to construct corresponding models
for Hot Star atmospheres is reviewed.

Since the physics of the atmospheres of Hot Stars are strongly
affected by velocity expansion dominating the spectra at all
wavelength ranges, hydrodynamic model atmospheres for O-type stars
and explosion models for Supernovae of Type~Ia are necessary as
basis for the synthesis and analysis of the spectra. It is shown
that stellar parameters, abundances, and stellar wind properties
can be determined by the methods of spectral diagnostics already
developed. Additionally, it will be demonstrated that models and
synthetic spectra of Type~Ia Supernovae of required quality are
already available. These will make it possible to tackle the question
of whether Supernovae~Ia are standard candles in a cosmological
sense, confirming or disproving that the current SN-luminosity
distances indicate accelerated expansion of the universe.

In detail we discuss applications of the diagnostic techniques by
example of two of the most luminous O~supergiants in the Galaxy and
a standard Supernova of Type~Ia. Furthermore, it is demonstrated
that the spectral energy distributions provided by state-of-the-art
models of massive O~stars lead to considerably better agreement
with observations if used for the analysis of H\,{\scriptsize
II}~regions. Thus, an excellent way of determining extragalactic
abundances and population histories is offered.

Moreover, the importance of Hot Stars in a broad astrophysical
context will be discussed. As they dominate the physical conditions
of their local environments and the life cycle of gas and dust
of their host galaxies, special emphasis will be given to the
corresponding diagnostic perspectives. Beyond that, the relevance
of Hot Stars to cosmological issues will be considered.}
\end{abstract}

\section{Introduction}

It is well known that Hot Stars are not a single group of
objects but comprise sub-groups of objects in different parts of
the HR~diagram and at different evolutionary stages. The most
important sub-groups are massive O/B stars, Central Stars of
Planetary Nebulae, and Supernovae of Type~Ia and~II. All these
sub-groups have in common that they are characterized by high
radiation energy densities and expanding atmospheres, and due
to this the state of the outermost parts of these objects is
characterized by non-equilibrium thermodynamics. In order to
cover the best-known fundamental stages of the evolution of Hot
Stars in sufficient depth this review will be restricted to the
discussion of O~stars and Supernovae of Type~Ia; we will not discuss
objects like Wolf-Rayet stars, Luminous Blue Variables, Be-stars,
Supernovae of Type~II, and others.  Furthermore, Hot Stars play an
important role in a broad astrophysical context. This implies that
a complete review covering all aspects in theory and observation is
not only impossible, but also beyond the scope of this review. Thus,
this review will focus on a special part of the overall topic with
the intent to concentrate on just one subject; the subject chosen
is UV spectral diagnostics. It will be shown, however, that this
subject has important implications for astrophysical topics which
are presently regarded as being ``trendy''.

But first of all we have to clarify what UV spectral diagnostics
means. This is best illustrated by the really old-fashioned (1977,
Morton and Underhill) UV~spectrum of one of the brightest massive
O~stars, the O4\,I(f) supergiant $\zeta$~Puppis. As can be seen
in Figure~1, expanding atmospheres have a pronounced effect on
the emergent spectra of hot stars -- especially in the UV-part.
The signatures of outflow are clearly recognized by the blue-shifted
absorption and red-shifted emission in the form of the well-known
P Cygni profiles. It is quite obvious that these kind of spectra
contain information not only about stellar and wind parameters,
but also about abundances. Thus, in principle, all fundamental
parameters of a hot star can be deduced from a comparison of observed
and synthetic spectra.

Although spectra of the quality shown in Figure~1 have been
available for more than 25 years, most of the work done to date has
concentrated on qualitative results and arguments. In view of the
effort put into the development of modern -- state-of-the-art --
instruments, it is certainly not sufficient to restrict the analyses
to simple line identifications and qualitative estimates of the
physical properties. The primary objective must be to extract the
complete physical stellar information from these spectra. Such
diagnostic issues principally have been made possible by the superb
quality and spectral resolution of the spectra available.

For this objective, the key is to produce realistic synthetic UV
spectra for Hot Stars. Such powerful tools, however, are still in
development and not yet widely used. They offer the opportunity
to determine the stellar parameters, the abundances of the light
elements -- He, C, N, O, Si -- and of the heavy elements like Fe
and Ni, quantitatively, as is indicated not only by IUE, but also
more recent HST, ORFEUS, and FUSE observations of hot stars in the
Galaxy and Local Group galaxies. All these observations show that
the spectra in the UV spectral range are dominated by a dense forest
of slightly wind-affected pseudo-photospheric metal absorption
lines\linebreak
% ----------------------------------------------------------------
\myfigure{\includegraphics[width=11.0cm]{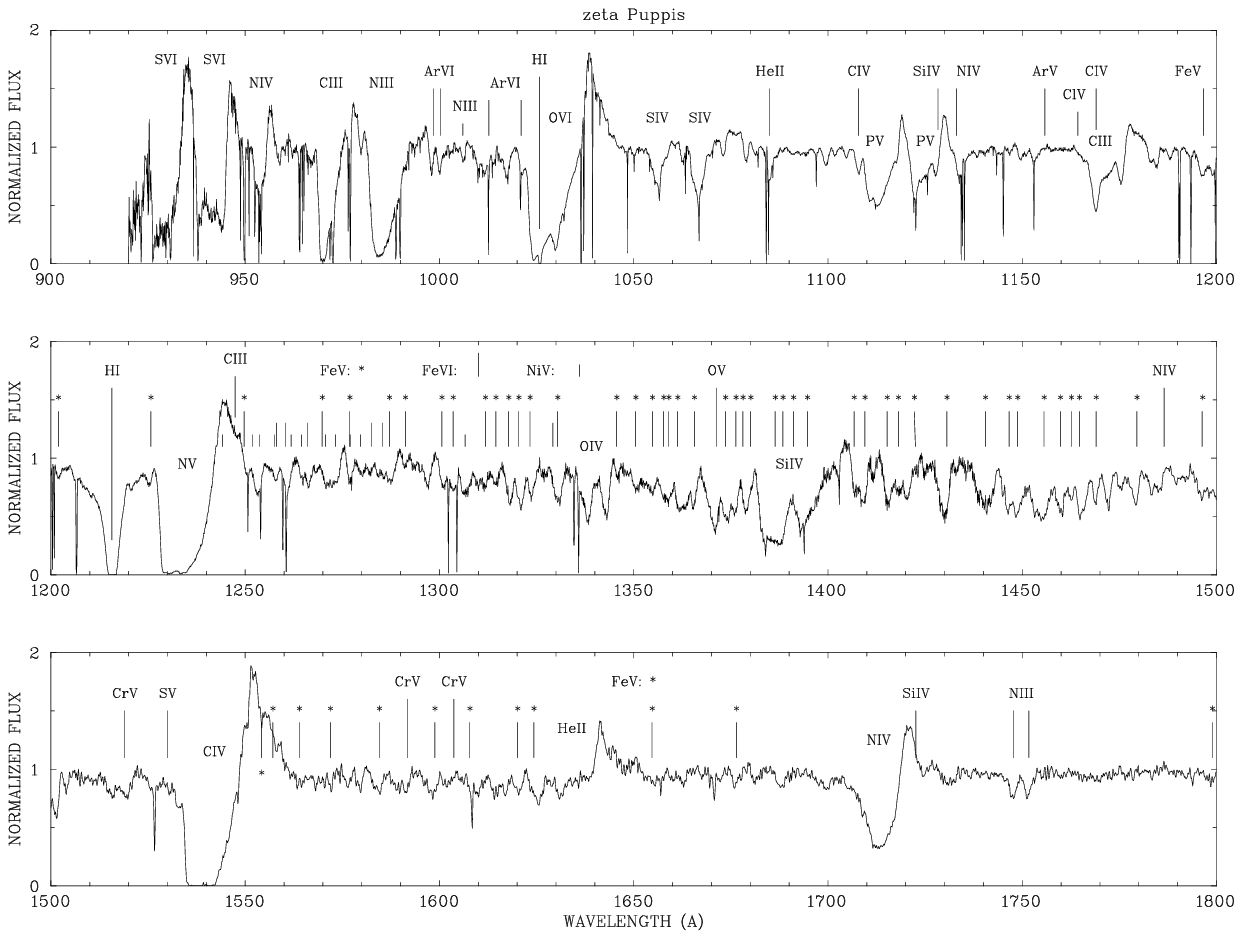}}
Figure 1: {\small Merged spectrum of
Copernicus and IUE UV high-resolution observations of the
O4\,I(f) supergiant $\zeta$ Puppis (900--1500~\AA: Morton and
Underhill~1977; 1500--1800~\AA: Walborn et\,al.\ 1985). The most
important wind lines of the light elements are identified and
marked. Also marked are the large number of wind-contaminated
lines of the iron group elements (e.g., \FeV) which are
especially present between 1250 and 1500~\AA. (Figure from
Pauldrach et\,al.\ 1994b).}\\
% ----------------------------------------------------------------

\noindent overlaid by broad P~Cygni line profiles of strong,
mostly resonance lines, formed in different parts of the expanding
atmosphere. Thus, the obvious objective is to investigate the
importance of these lines with respect to the structure of the
expanding atmospheres that are characterized by the strength and
the velocity of the outflow, and through which the shape of the
spectral lines is mainly determined.  Over the last 30 years it
turned out that the achievement of this objective, which will also
be the primary aim of this paper to review, remains a difficult task.

Before we discuss in detail the status quo of the diagnostic tool
required (Section~6), we will first examine whether such a tool has
been made obsolete by the general development in astrophysics or
whether it is still relevant to current astronomical research.

For this purpose we first discuss the diagnostic perspectives of
galaxies with pronounced current star formation. Due to the impact
of massive stars on their environment the physics underlying
the spectral appearance of starburst galaxies are rooted in the
atmospheric expansion of massive O~stars which dominate the UV
wavelength range in star-forming galaxies. Therefore, the
UV-spectral features of massive O~stars can be used as tracers of
age and chemical composition of starburst galaxies even at high
redshift (Section~2).

With respect to the present cosmological question of the
reionization of the universe -- which appeared to have happened at
a redshift of about $z \sim 6$ -- the ionization efficiency of a
top-heavy Initial Mass Function for the first generations of stars
is discussed (Section~3).

Starting from the impact of massive stars on their environment it
is demonstrated that the spectral energy distributions provided by
state-of-the-art models of massive O~stars lead to considerable
improvements if used for the analysis of
\HII~regions. Thus, the corresponding methods for determining
extragalactic abundances and population histories are promising
(Section~4).

Regarding diagnostic issues, the role of Supernovae of Type~Ia as
distance indicators is discussed. The context of this discussion
concerns the current and rather surprising surprising result that
distant SNe~Ia appear
fainter than standard candles in an empty Friedmann model of the
universe (Section~5).

Finally, we discuss applications of the diagnostic techniques by
example of two of the most luminous O~supergiants in the Galaxy;
additionally, basic steps towards realistic synthetic spectra for
Supernovae of Type~Ia are presented (Section~7).

\section{The impact of massive stars on their Environment --\\
                UV Spectral Analysis of Starburst Galaxies}

The impact of massive stars on their environment in the present
phase of the universe is of major importance for the evolution of
most galaxies. Although rare by number, massive stars dominate the
life cycle of gas and dust in star forming regions and are
responsible for the chemical enrichment of the ISM, which in turn
has a significant impact on the chemical evolution of the host
galaxy. This is mainly due to the short lifetimes of massive stars,
which favours the recycling of heavy elements in an extremely
efficient way. Furthermore, the large amount of momentum and energy
input of these objects into the ISM controls the dynamical
evolution of the ISM. This takes place in an extreme way, because
massive stars mostly group in young clusters, producing void
regions around themselves and wind- and supernova-blown
superbubbles around the clusters. These superbubbles are ideal
places for further star formation, as numerous Hubble Space
Telescope images show. Investigation of these superbubbles will
finally yield the required information to understand the various
processes leading to continuous star formation regions (cf.~Oey and
Massey 1995). The creation of superbubbles is also responsible for
the phenomenon of galactic energetic outflows observed in
starbursts (Kunth et\,al.\ 1998) and starburst galaxies even at
high redshift (Pettini et\,al.\ 1998). It is thus not surprising
that spectroscopic studies of galaxies with pronounced current star
formation reveal the specific spectral signatures of massive stars,
demonstrating in this way that the underlying physics for the
spectral appearance of starburst galaxies is not only rooted in the
atmospheric expansion of massive O~stars, but also dominated by
these objects (cf.~Figure~2 from Steidel et\,al.\ 1996, for
star-forming galaxies at high redshift see also Pettini et\,al.\
2000, and for UV line spectra of local star-forming galaxies see
Conti et\,al.\ 1996; note that the similarity of the spectra at
none/low and high redshifts suggests a similar stellar content).

% ----------------------------------------------------------------
\myfigure{\includegraphics[width=9.9cm]{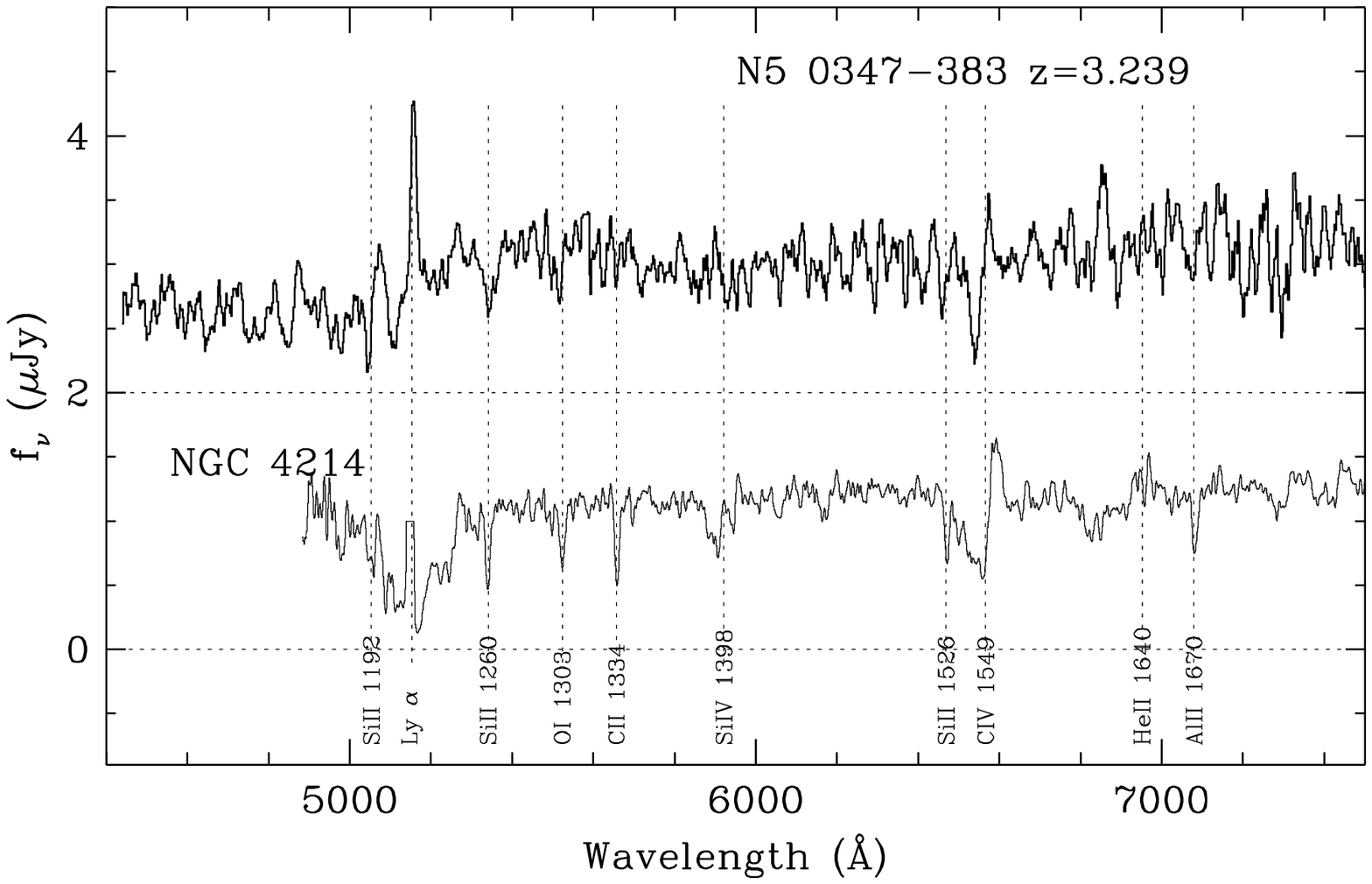}}
Figure 2: {\small UV spectrum of a $z > 3$ galaxy (upper part). For
comparison, a recent HST spectrum of the central starburst region
in the Wolf-Rayet galaxy NGC~4214 is also shown (lower part). Note
the characteristic P-Cygni lines, especially of \CIV and \SiIV,
pointing to the dominating influence of massive O~stars. Figure
from Steidel et\,al.\ 1996.}\\
% ----------------------------------------------------------------

The characteristic P-Cygni lines observed as broad stellar wind
lines, especially those of the resonance lines of \CIV\ and
\SiIV, integrated over the stellar populations in the spectra of
starbursting galaxies, allow quantitative spectroscopic studies
of the most luminous stellar objects in distant galaxies even at
high redshift. Thus, in principle, we are able to obtain important
quantitative information about the host galaxies of these objects,
but diagnostic issues of these spectra require among other things
{\it synthetic UV spectra of O-type stars\/} as input for the
population synthesis calculations needed for a comparison with the
observed integrated spectra.

The potential of these spectra for astrophysical diagnostics
can nevertheless be investigated in a first step by using {\it
observed\/} UV spectra of nearby O-type stars as input for the
corresponding population synthesis calculations instead. In the frame
of this method stars are simulated to form according to a specified
star-formation history and initial mass function and then follow
predefined tracks in the HR-diagram. The integrated spectra are
then built up from a library of observed UV~spectra of hot stars in
the Galaxy and the Magellanic Clouds. The output of this procedure
are semi-empirical UV spectra between 1000 and 1800~\AA\ at 0.1 to
0.7~\AA\ resolution for populations of arbitrary age, star-formation
histories, and initial mass function. The computational technique
of this method is described comprehensively in the literature
and we refer the reader to one of the latest papers of a series
(Leitherer et\,al.\ 2001).

As an example of the analyses performed in this way a comparison
of the average spectrum of 8 clusters in NGC~5253 to synthetic
models at solar (top) and 1/4~solar (bottom) metallicity is shown
in Figure~3 (Leitherer et\,al.\, 2001). The Figure shows clearly
that a representative value of the overall metallicity of this
starburst galaxy can be determined, since the model spectrum in
the lower part fits the observation\linebreak
% ----------------------------------------------------------------
\myfigure{\includegraphics[angle=90,width=10.2cm]{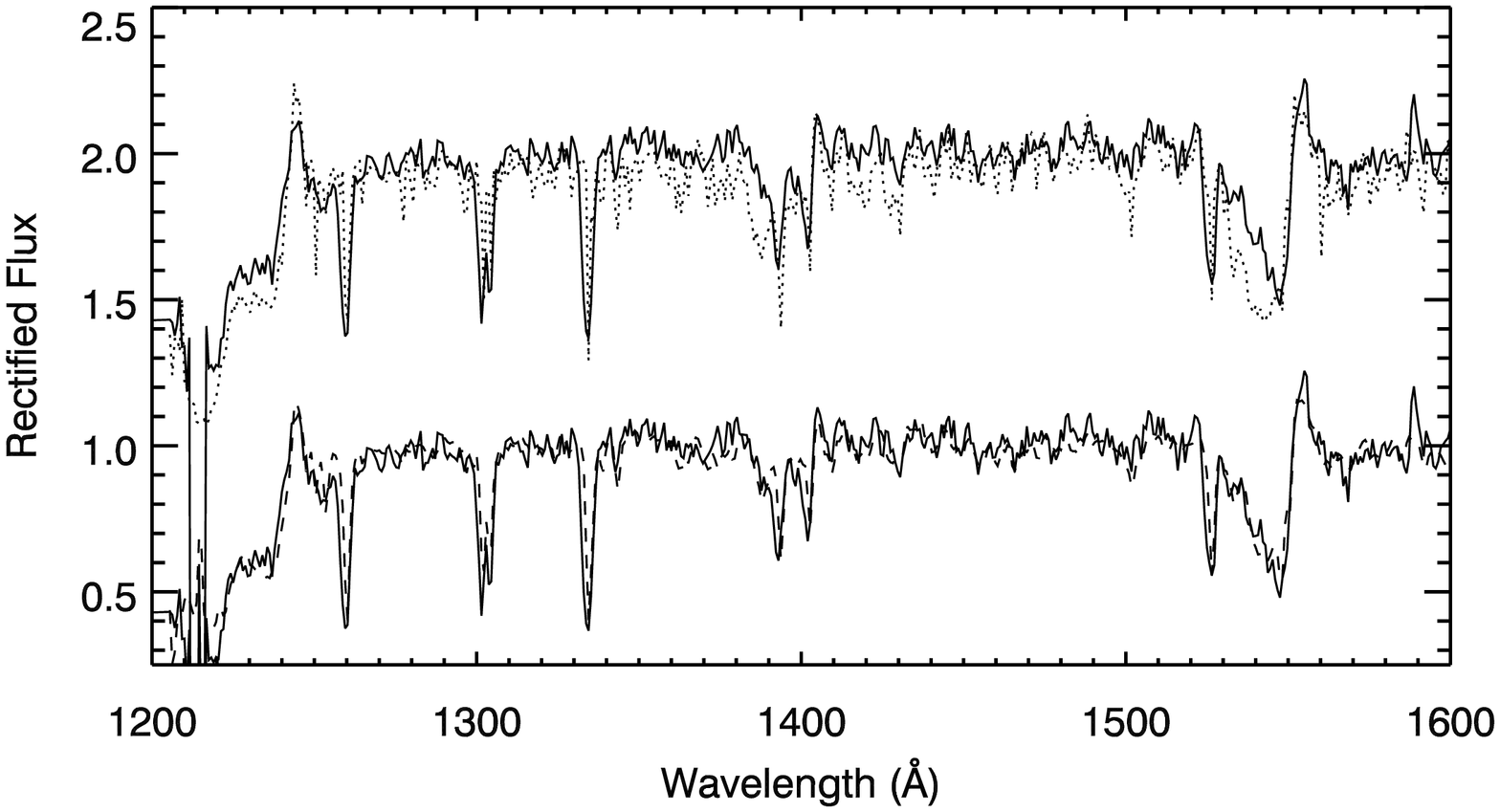}}
Figure 3: {\small Average spectrum of 8 clusters in NGC~5253
compared to that of a population model at solar (top) and 1/4~solar
metallicity (bottom). It is clearly shown that a representative
value of the overall metallicity of this starburst galaxy can be
determined. Figure from Leitherer et\,al., 2001.}\\
% ----------------------------------------------------------------

\noindent almost perfectly. This result is especially convincing
as both models (dashed lines) are based on the same parameters,
except for the metallicity.  A standard Salpeter IMF between 1 and
100~$\Msun$ was used and the starburst has been assumed to last
6~Myr, with stars forming continuously during this time. The worse
fit to the observations produced by the solar metallicity model
spectrum -- particularly discrepant are the blue absorption wings
in \SiIV\ and \CIV\ which are too strong in the models -- could be
improved by reducing the number of the most massive stars with a
steeper IMF, but at the cost of the fit quality in the emission
components. As an important result the ratio of the absorption
to the emission strengths is therefore a sensitive indicator for
the metallicity.

Moreover, the strong sensitivity of the emission parts of the
P-Cygni lines of \NV, \SiIV, and \CIV\ on the evolutionary stage of
the O~stars makes these spectra quite suitable as age tracers,
which is shown in Figure~4, where the time evolution of the integrated
spectrum following an instantaneous starburst is presented. The
line profiles gradually strengthen from a main-sequence-dominated
population at 0--1~Myr to a population with luminous O~supergiants
at 3--5~Myr; and the lines weaken again at later age due to
termination of the supergiant phase -- this behavior is especially
visible in the shape of the \CIV\ $\lambda 1550$ resonance line.

The conclusion from these examples is that this kind of analysis is
very promising, but relies on observed UV~spectra, which are just
available for a small number of metallicity values, namely those of
the Galaxy and the Magellanic Clouds. In order to make progress in
the direction outlined before, {\it realistic synthetic UV Spectra
of O-type stars\/} are needed. This is in particular the case for
high-redshift galaxies (which are observable spectroscopically when the flux
is amplified by gravitational lensing through foreground galaxy
clusters) since in these cases the expected metallicities of
starbursting galaxies in the early universe (cf.~Pettini et
al.~2000) are most probably different from local ones.

% ----------------------------------------------------------------
\myfigure{\includegraphics[angle=90,width=10.0cm]{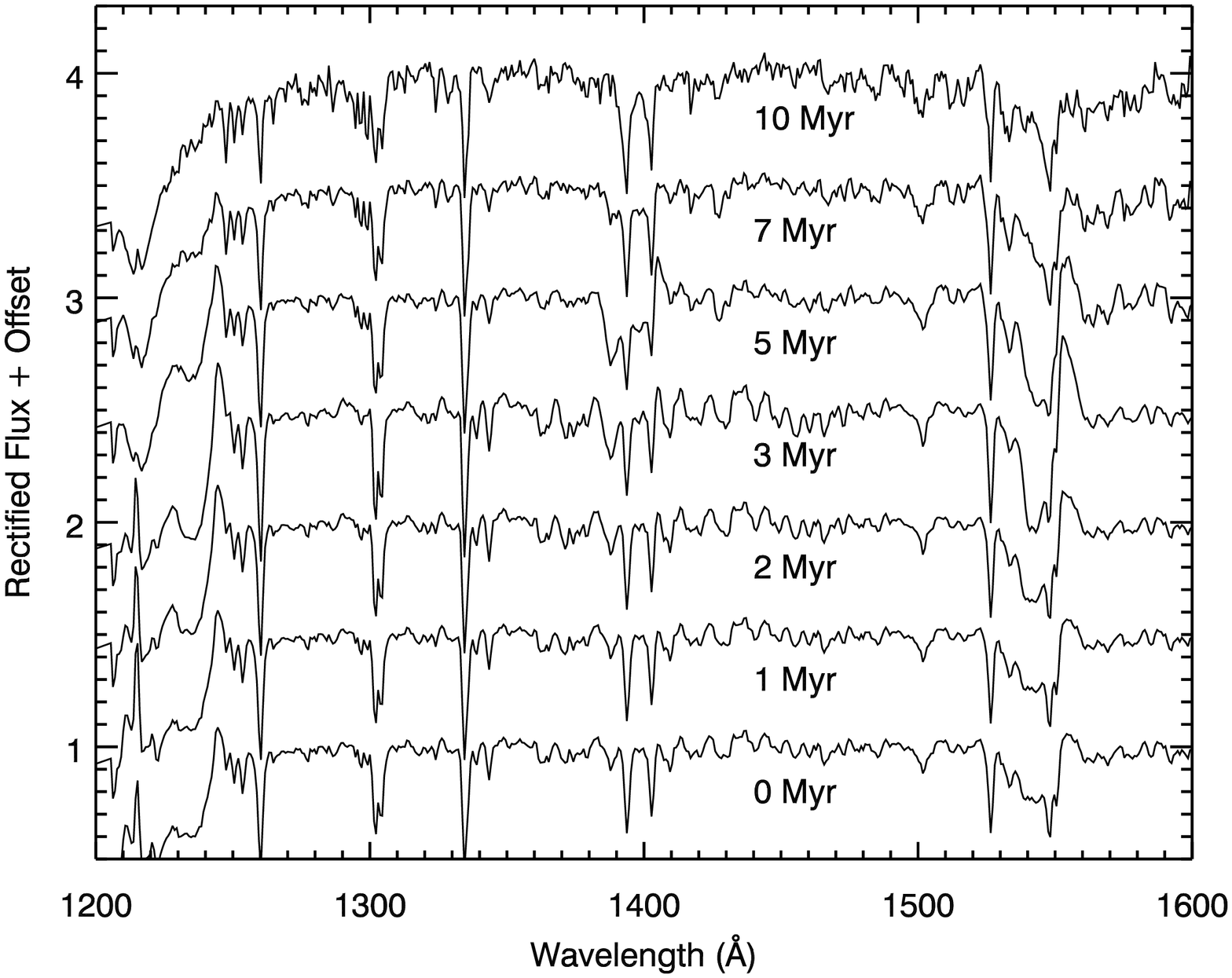}}
Figure 4: {\small Time evolution of the integrated spectrum for the
10~Myr following an instantaneous starburst. The strong sensitivity
of the emission parts of the P-Cygni lines of \NV, \SiIV, and \CIV\
on the evolutionary stage of the O~stars makes these spectra quite
suitable as age tracers. Figure from Leitherer et\,al.\ 2001.}\\
% ----------------------------------------------------------------

\section{First generations of Stars --
Ionization Efficiency\\ of a Top-Heavy Initial Mass Function}

Apart from the short evolutionary timescale of massive stars it is
obviously the metallicity, and, connected to this, the steepness of
the Initial Mass Function, which is responsible for the rarity of
these objects in the present phase of the universe. It has to be
the metallicity, because very recently strong evidence has been
found that the primordial IMF has favored massive stars with masses
$>10^2 \Msun$ (cf.~Bromm et\,al.\ 1999). Thus, in the early
universe, when only primordial elements were left over from the Big
Bang, nature obviously preferred to form massive stars. This
prediction is based on the missing metallicity which leads to a
characteristic scale for the density and temperature of the
primordial gas, which in turn leads to a characteristic Jeans mass
of $M_{\rm J}\sim 10^3 \Msun$ (Larson~1998).

Finally, due to these physical conditions the Initial Mass Function
becomes top-heavy and therefore deviates significantly from the
standard Salpeter power-law (see, for instance, Bromm et\,al.\
1999, 2001). Such an early population of very massive stars at
very low metallicities (i.\,e., Population~III stars) which have
already been theoretically investigated by El Eid et\,al.\ (1983),
recently turned out to be also relevant to cosmological issues,
the most important being the cosmological question of when\linebreak
% ----------------------------------------------------------------
\myfigure{\includegraphics[width=8.4cm]{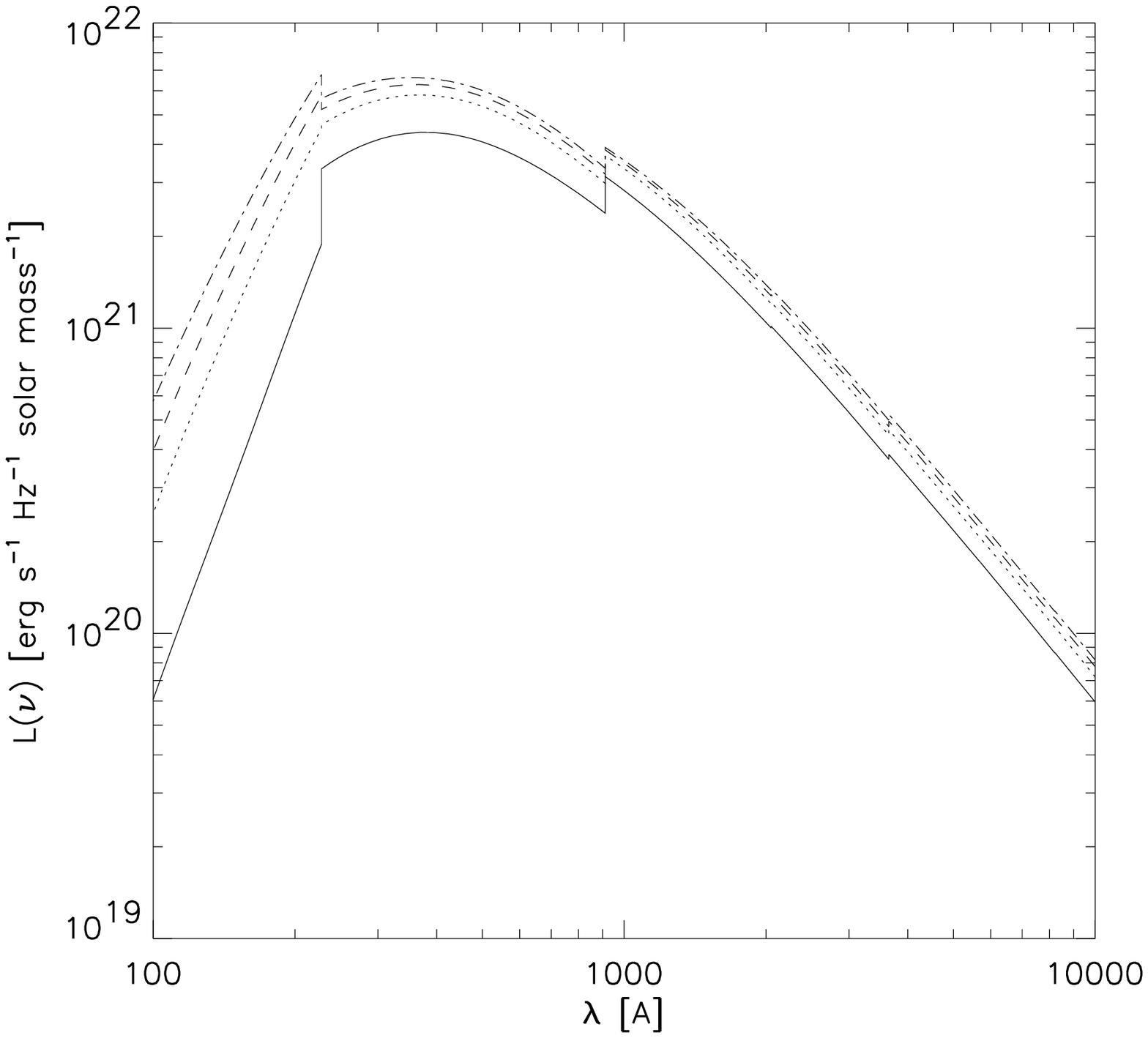}}
Figure 5: {\small The normalized spectral
energy distribution in the continuum of Population III stars --
mass range 100--1000~$\Msun$ at $Z=0$. Note that the spectra
attain an universal form for stellar masses $M > 300~\Msun$
Figure from Bromm et\,al.\ 2001.}\\
% ----------------------------------------------------------------

\noindent and how the cosmic ``dark ages'' ended (Loeb 1998). We
know that the dark ages ended because the absence of a Gunn-Peterson
trough (Gunn and Peterson 1965) in the spectra of high-redshift
quasars implies that the universe was reionized again at a redshift
of $z> 5.8$ (Fan et\,al.\ 2000). For a long time, Carr et\,al.\
(1984) suspected that the first generations of stars have been
relevant to control this process. Thus, Population III stars
could contribute significantly to the ionization history of the
intergalactic medium (IGM), but the contribution of the first
generation of stars to the ionization history of the IGM depends
crucially on their initial mass function. With regard to the first
generations of stars the ionization efficiency of a top-heavy
Initial Mass Function will, therefore, have to be investigated.

It is the enormous amount of UV and EUV radiation of these very
massive stars which could easily change the status of the cold and
dark universe at that time to become reionized again. This is
indicated in Figure~5, which also shows that the total spectral
luminosity depends solely on the total amount of mass, if the mass
of the most massive stars exceeds 300~$\Msun$ (cf.~Bromm et\,al.\
2001). Due to this top-heavy Initial Mass Function the total
spectral energy distribution deviates significantly from that
obtained with the standard Salpeter power-law, as is shown in
Figure~6, and the flux obtained can contribute the decisive part to
the unexplained deficit of ionizing photons required for the
reionization of the universe (cf.~Bromm et\,al.\ 2001).

However, in order to be able to make quantitative predictions
about the influence of this extremely metal poor population of very
massive stars on their galactic and intergalactic environment one
primarily needs observations that can be compared to the predicted
flux spectrum.

% ----------------------------------------------------------------
\myfigure{\includegraphics[width=7.9cm]{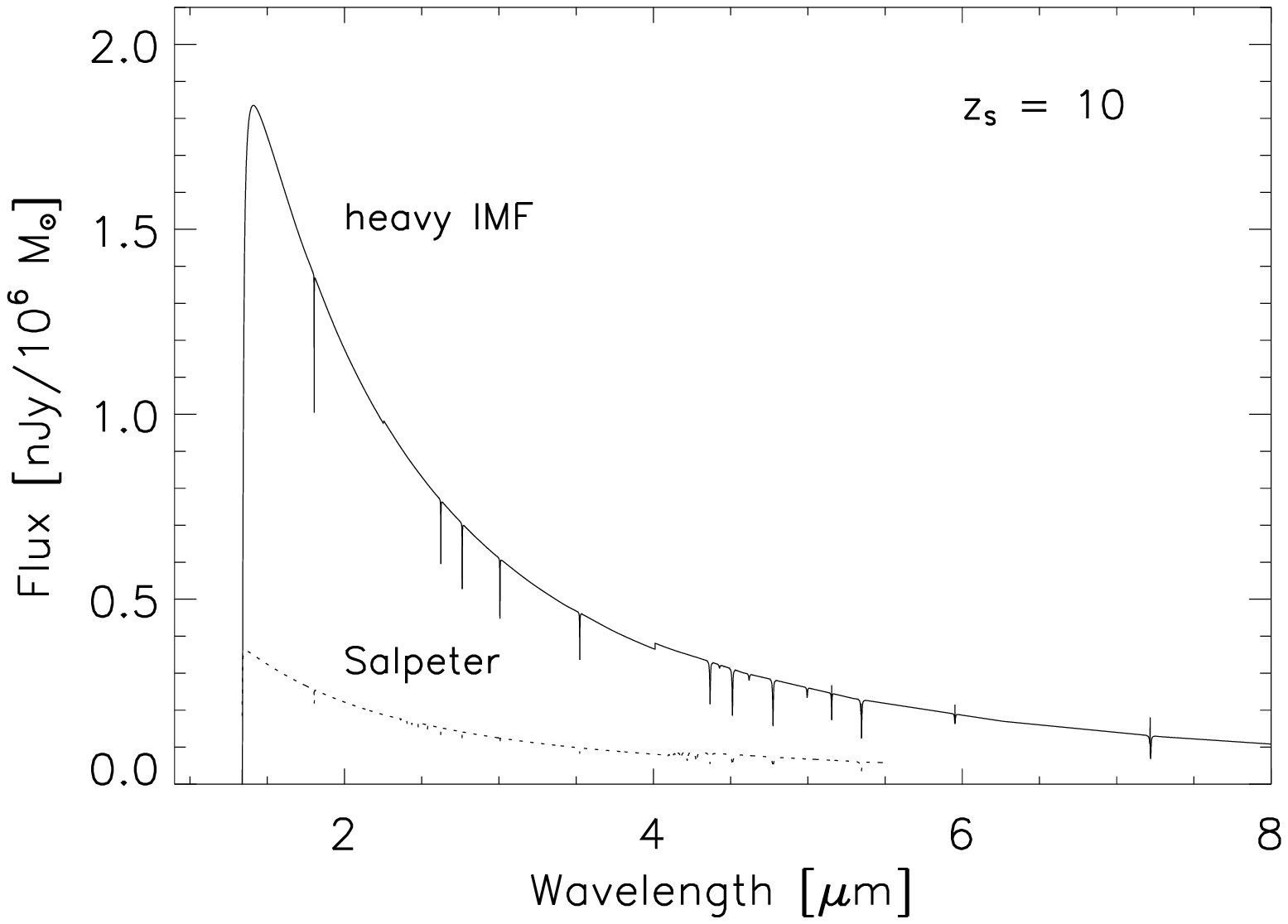}}
Figure 6: {\small Predicted flux from a Population~III star cluster
at $z=10$. A flat universe with $\Omega_{\Lambda}= 0.7$ is assumed.
The cutoff below $\lambda_{\rm obs} = 1216$~\AA\ is due to complete
Gunn-Peterson absorption, and the observable flux is larger by an
order of magnitude for the case of the top-heavy IMF when compared
to the case of the standard Salpeter power-law. Figure from Bromm
et\,al.\ 2001.}\\
% ----------------------------------------------------------------

Future observations with the Next Generation Space Telescope of
distant stellar populations at high redshifts will principally
give us the opportunity to deduce the primordial IMF from these
comparisons.  Kudritzki~(2002) has recently shown that this is
generally possible, by calculating state-of-the-art UV~spectra for
massive O~stars in a metallicity range of $1\dots 10^{-4}~\Zsun$.
From an inspection of his spectra he concluded that significant line
features are still detectable even at very low metallicities; thus,
there will be diagnostic information available to deduce physical
properties from starbursting regions at high redshifts that will
eventually be observed with the Next Generation Space Telescope in
the infrared spectral region.

As a second requirement we need to determine the physical properties
of Population~III stars during their evolution. A key issue in this
regard is to obtain {\it realistic spectral energy distributions\/}
calculated for metallicities different from zero for the most
massive objects, since the assumption of a metallicity of $Z = 0$
is certainly only correct for the very first generation of
Population~III stars.

\section{Theoretical Ionizing Fluxes of O~Stars}

Although less spectacular, we will now investigate the impact of
massive stars on their environment in a more direct manner. Apart
from the chemical enrichment of the ISM, the large amount of
momentum and energy input into the ambient interstellar medium of
these objects is primarily of importance. Especially the radiative
energy input shortward of the Lyman~edge, which ionizes and heats
the Gaseous Nebulae surrounding massive Hot Stars, offers the
possibility to analyze the influence of the EUV~radiation of the
photoionizing stars on the ionization structure of these excited
\HII~regions.

% ----------------------------------------------------------------
\myfigure{\includegraphics[angle=90,width=9.5cm]{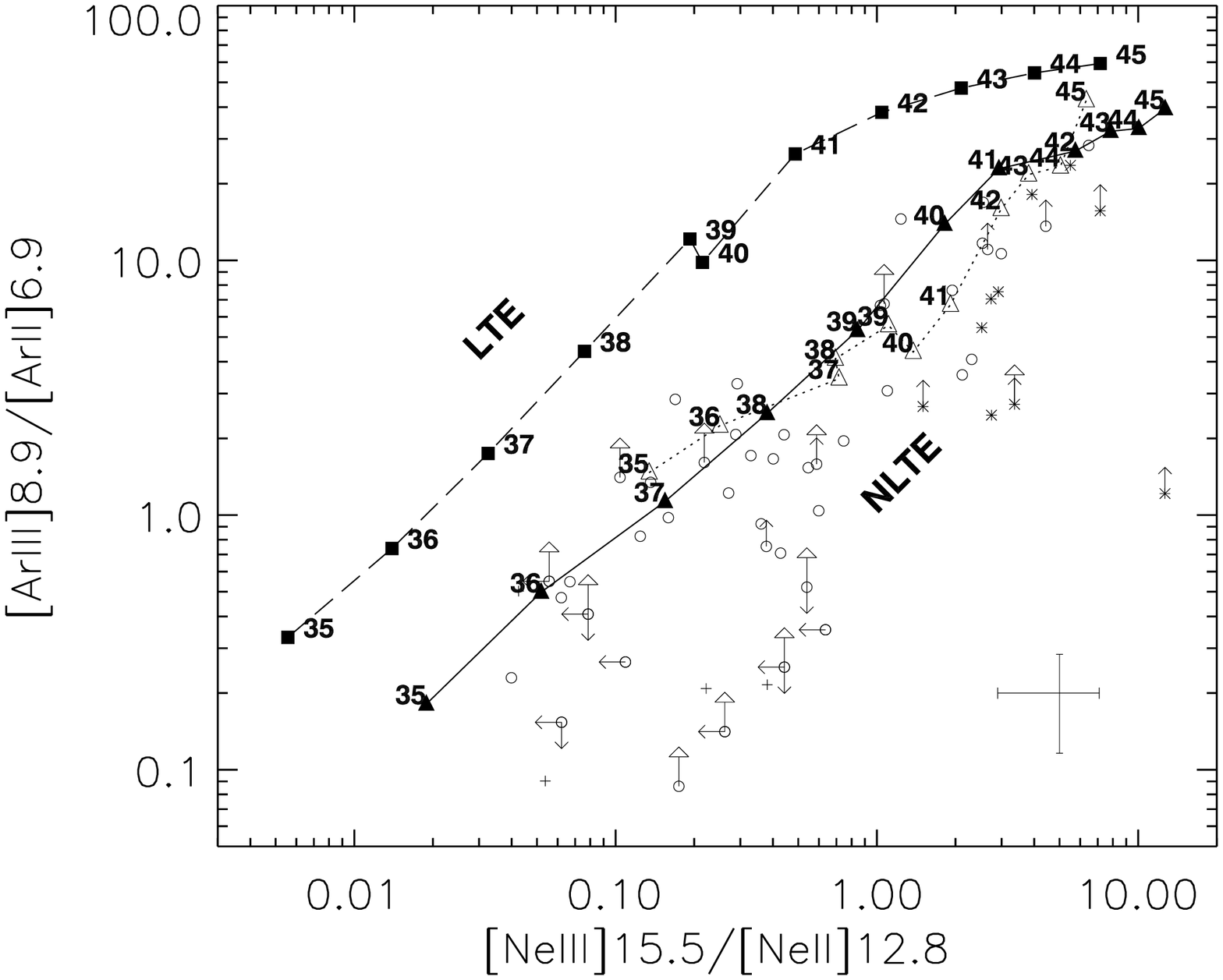}}
Figure 7: {\small [\ArIII]
8.99~$\mu$m\,/\,[\ArII] 6.99~$\mu$m versus [\NeIII]
15.6~$\mu$m\,/\,[\NeII] 12.8~$\mu$m diagnostic
diagram of observed and predicted nebular excitation
of \HII~regions. Boldface numbers indicate models
which are designated by their effective temperature in
$10^3$\,K. Triangles are NLTE models and squares are LTE models;
note that the NLTE models represent an impressive improvement
in reproducing the [\NeIII] emission.  A representative error bar
for the data is shown in the lower right corner.  Figure from
Giveon et\,al.\ 2002.}\\
% ----------------------------------------------------------------

The primary objective of such investigations are studies of
theoretical models of starburst regions, which for instance are
used to determine the energy source in ultra-luminous infrared
galaxies -- ULIRGs -- (cf.~Lutz et\,al.\ 1996; Genzel et\,al.\
1998). The interpretation of the corresponding extra-galactic
observations obviously requires understanding the properties of the
spectral energy distributions (SEDs) of massive stars and stellar
clusters. Thus, the quality of the SEDs has to be probed in a first
step by means of investigations of Galactic \HII~regions.

\subsection{The Excitation of Galactic H\,{\small II}~Regions}

Giveon et\,al.\ (2002) recently presented a comparison of
observed [\NeIII] 15.6~$\mu$m\,/ [\NeII] 12.8~$\mu$m and [\ArIII]
8.99~$\mu$m\,/\,[Ar II] 6.99~$\mu$m excitation ratios, obtained for
a sample of 112 Galactic \HII~regions and 37 nearby extragalactic
\HII~regions in the LMC, SMC, and M33 observed with ISO-SWS, with
the corresponding results of theoretical nebular models.

The authors have chosen infrared fine-structure emission lines for
their investigation because these lines do not suffer much from
dust extinction and the low energies of the associated levels make
these lines quite insensitive to the nebular electron temperature.
Moreover, the relative strengths of the fine-structure emission
lines chosen are ideal for constraining the shape of the
theoretical ionizing fluxes,\linebreak
% ----------------------------------------------------------------
\myfigure{\includegraphics[width=11.0cm]{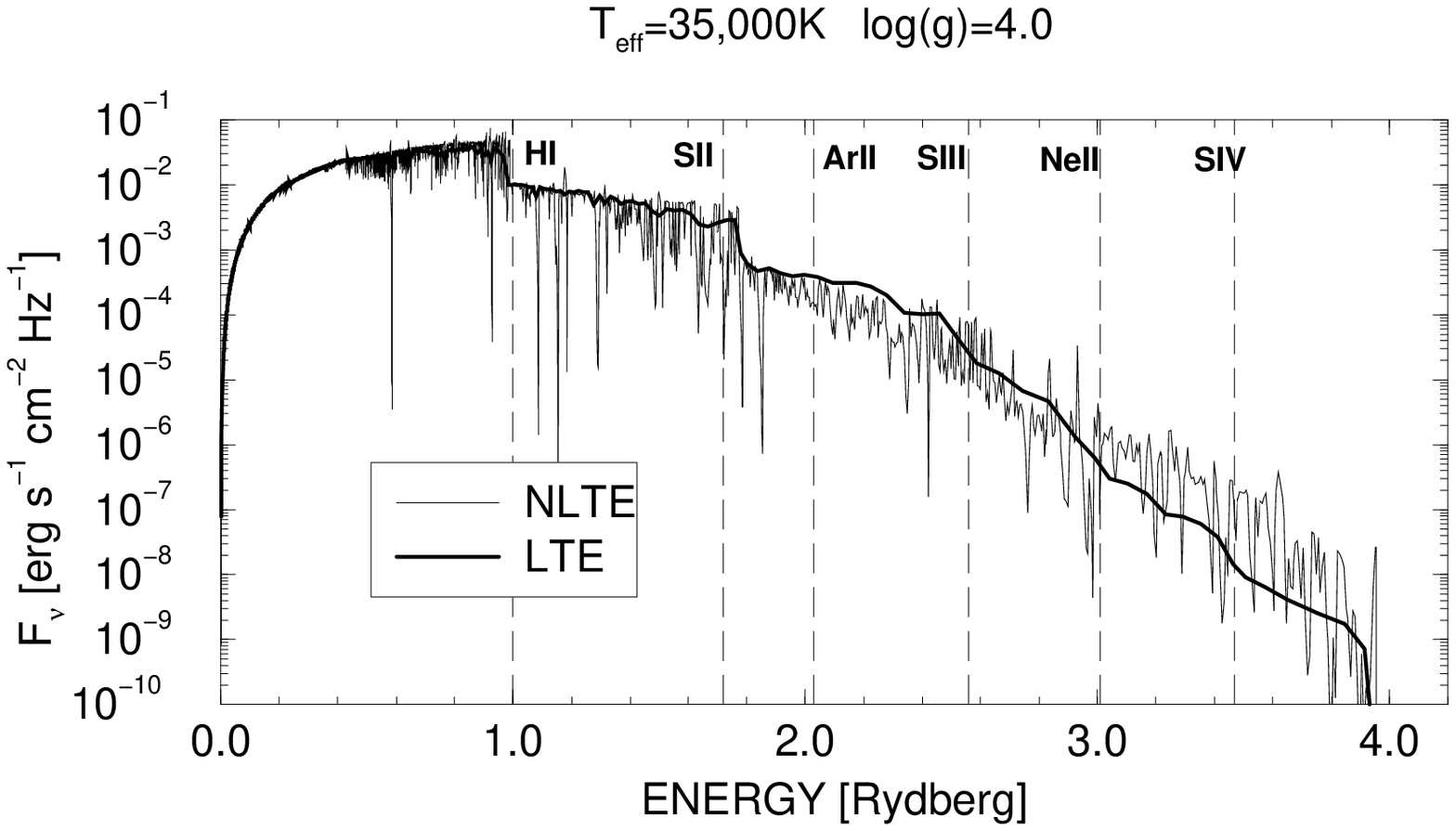}}
Figure 8: {\small Comparison between the LTE (static atmosphere
from Kurucz~1992) and NLTE (expanding atmosphere from Pauldrach
et\,al.\ 2001) spectral energy distributions of a dwarf with an
effective temperature of 35\,000~K. The ionization edges of the
relevant ions are indicated by vertical dashed lines. It is
apparent that the difference in excitation produced by the LTE and
NLTE atmospheres will be more pronounced in the Ne$^{++}$/Ne$^+$
ratio than in the Ar$^{++}$/Ar$^+$ ratio. Figure from Giveon
et\,al.\ 2002.}\\
% ----------------------------------------------------------------

\noindent since the [\NeIII] 15.6~$\mu$m\,/\,[\NeII] 12.8~$\mu$m
line ratio depends on photons emitted at $\geq 3$~Rydbergs, while
[\ArIII] 8.99~$\mu$m\,/\,[\ArII] 6.99~$\mu$m is sensitive to the
region $\geq 2$~Rydbergs. Thus, these line ratios are extremely
useful probes of the physical properties of \HII~regions and their
associated ionizing sources especially with regard to the ionizing
spectral energy distributions.

The complete set of observed emission line ratios of Ne and Ar is
shown in Figure~7 together with the corresponding results of nebular
model computations for which solar metallicity and a gas density of
800~cm$^{-3}$ has been assumed, and which are based on LTE (static
atmosphere from Kurucz,~1992) as well as NLTE (expanding atmosphere
from Pauldrach et\,al., 2001) spectral energy distributions.

The diagnostic diagram clearly shows that the predicted nebular
excitation increases with increasing effective temperature of the
photoionizing star, and it also shows that the high excitation
[\NeIII] emission observed in \HII~regions is by far not reproduced
by nebular calculations which make use of the ionizing fluxes of
LTE models -- the line ratios are under-predicted by factors larger
than 10. This result is clearly an example of the well-known \NeIII\
problem (cf.~Baldwin et\,al.\ 1991; Rubin et\,al.\ 1991; Simpson
et al.~1995).

It is quite evident, however, that the quality of the diagnostics
depends primarily on the quality of the input, i.\,e., the spectral
energy distributions. It is therefore a significant step forward
that the ionizing fluxes of the NLTE models used represent an
impressive improvement to the observed excitation correlation --
note that the fit shown in Figure~7 may actually be even better,
since a lot of the scatter is due to underestimated extinction
corrections for the [\ArIII] 8.99~$\mu$m line. The NLTE sequence
also indicates that for most of the considered \HII~regions the
effective temperatures of the exciting stars lie in the range of
35\,000 to 45\,000~K. Most importantly, the NLTE models can readily
account for the presence of high excitation [\NeIII] emission lines
in nebular spectra. This result resolves the long-standing \NeIII\
problem and supports the conclusion of Sellmaier et\,al.\ (1996),
who found, for the first time, on the basis of a less comprehensive
sample of \HII~regions and models that this problem has been the
failure of LTE photoionization simulations primarily due to a
significant under-prediction of Lyman photons above 40~eV. This
issue is solved by making use of NLTE model atmospheres.

The reason for this improvement are the spectral shapes of the NLTE
fluxes shortward of the \ArII\ and \NeII\ ionization thresholds
which are obviously somewhat more realistic in the NLTE case. This
is illustrated in Figure~8 where the spectral energy distributions
of an LTE and an NLTE model are compared by example of a dwarf with an
effective temperature of 35\,000~K. As is shown, the SEDs harden in
the NLTE case, meaning that the LTE model produces much less flux
above the Ne$^+$ 40.96~eV threshold than the NLTE model. This is
the behavior that is essentially represented in the excitation
diagram.

\subsection{Spectral Energy Distributions\\ of Time-Evolving
Stellar Clusters}

The improvement obtained for the diagnostic diagram of the analysis
of Galactic \HII~regions in Section~4.1 also has important
implications for determining extragalactic abundances and
population histories of starburst galaxies.

With regard to this, Thornley et\,al.\ (2000) carried out detailed
starburst modelling of the [\NeIII] 15.6~$\mu$m\,/\,[\NeII]
12.8~$\mu$m ratio of \HII~regions ionized by clusters of stars. As
was shown above, the hottest stars in such models are responsible for
producing large nebular [\NeIII] 15.6~$\mu$m\,/\,[\NeII] 12.8~$\mu$m
ratios. Hence, the low ratios actually observed led to the conclusion
that the relative number of hot stars is small due to aging of the
starburst systems. Thus, the solution of the \NeIII\ problem has
important consequences for the interpretation of these extragalactic
fine-structure line observations, since the conclusion that due to
the low [\NeIII] 15.6~$\mu$m\,/\,[\NeII] 12.8~$\mu$m ratios obtained,
the hottest stars as dominant contributors to the ionization of the
starburst galaxies have to be removed, has obviously to be proven.

In order to tackle this challenge, Sternberg et\,al.\ (2002)
computed spectral energy distributions of time-evolving stellar
clusters, on the basis of a large grid of calculated NLTE spectral
energy distributions of O and early B-type stars
(being the major contributors to the Lyman and \HeI\ fluxes)
in the hot, luminous part ($\Teff >$ 25\,000~K) of the HR diagram.
From this grid, models of stars following evolutionary tracks are suitably
interpolated. The SEDs used rely on recent improvements of
modelling expanding NLTE atmospheres of Hot Stars (Pauldrach
et\,al.\ 2001). As an example of the individual models used,
Figure~9 shows the calculated spectral energy distribution of a
typical O~star compared to the corresponding result of an NLTE
model of Schaerer and de Koter (1997).

% ----------------------------------------------------------------
\myfigure{\includegraphics[width=9.5cm]{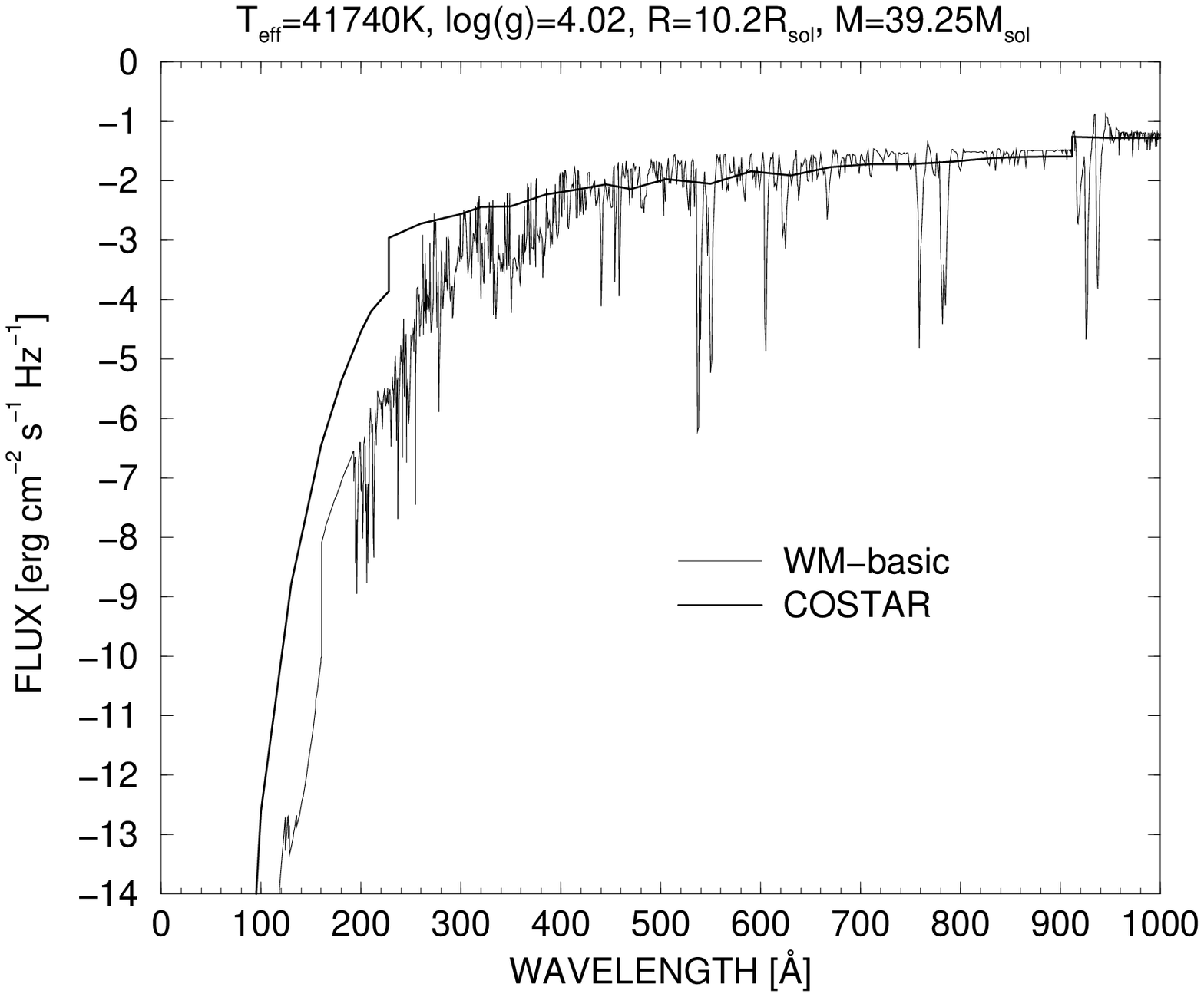}}
Figure 9: {\small Calculated spectral energy distribution of a
typical O~star. The spectrum represents the result of the WM-basic
model (Pauldrach et\,al.\ 2001), and the thick line represents the
result of a corresponding COSTAR model (Schaerer and de
Koter~1997).}\\
% ----------------------------------------------------------------

The spectral energy distributions have been calculated for two
modes of cluster evolution, a {\it continuous\/} and an {\it
impulsive\/} one. The different spectral evolutions of these modes
are shown in Figure~10 with regard to the spectral range of
0.8--2.5 Rydbergs.

In the continuous mode, the star-formation rate is assumed to be
constant with time -- on time scales which are longer than the
lifetimes of massive stars -- and the cluster is assumed to form
stars at a rate of 1~$\Msun$ per year. In the impulsive mode, all
stars are formed ``instantaneously'', i.\,e., on time scales which
are much shorter than the lifetimes of massive stars; the total
mass of stars formed is $10^5\,\Msun$, and the cluster is assumed
to evolve passively thereafter. Furthermore, the computations are
based on a Salpeter initial mass function, where for each mode two
cases are assumed, one with an upper mass limit of $M_{\rm up} =
120~\Msun$ and the other one with an upper mass limit of $M_{\rm up}
= 30~\Msun$.

\medskip \noindent
Two striking effects are seen in Figure~10:

\smallskip \noindent
1. In the continuous mode, the cluster ionizing spectrum becomes
softer -- i.e., steeper -- between 1 and 10~Myr. This is due to the
increase of the relative number of late versus early-type OB stars
during the evolution, since the less massive late-type stars have
longer lifetimes. This effect is particularly noticeable for the
cluster model with $M_{\rm up} = 120~\Msun$.

\smallskip \noindent
2. In the impulsive mode, the magnitude of the Lyman break
increases drastically because the massive stars disappear with
time.

% ----------------------------------------------------------------
\myfigure{\includegraphics[angle=90,width=\textwidth]{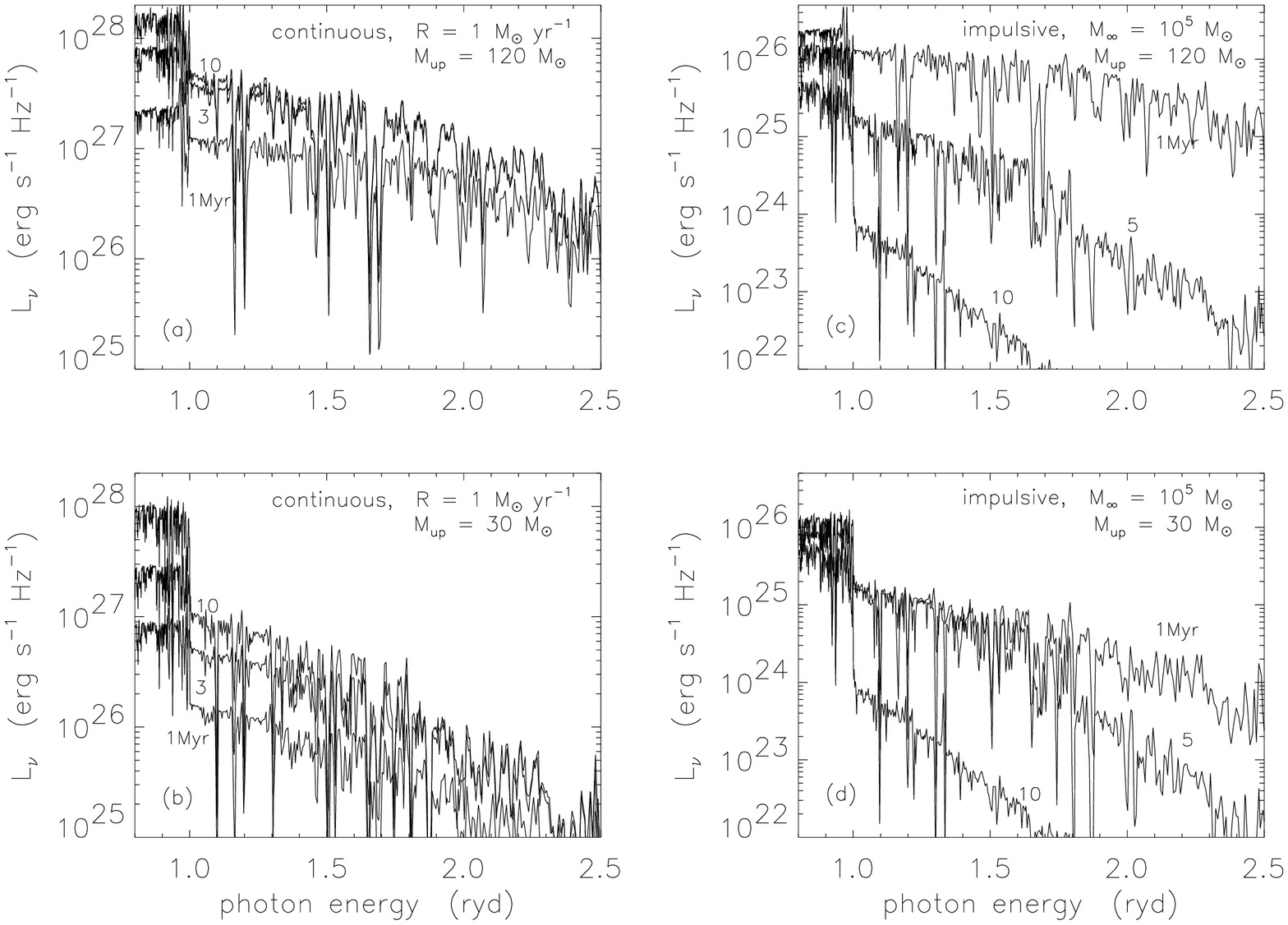}}
Figure 10: {\small Spectral energy distributions of time-evolving
stellar clusters. Two modes of cluster evolution are shown: {\it
continuous\/} and {\it impulsive\/}. {\em (Left)\/}: Evolution for
continuous star formation (lasting $10^{10}$\,yr) for two extreme
stellar compositions. In the upper panel evolving SEDs of a cluster
with $M_{\rm up} = 120~\Msun$, and in the lower panel with $M_{\rm
up} = 30~\Msun$ are shown. Note that at early phases the amount of
both ionizing and non-ionizing photons increase due to the
increasing number of new-formed stars. After a time period of $\sim
5$~Myr the number of ionizing photons remains roughly constant,
because an equilibrium of stellar aging and stellar birth is
achieved, whereas the number of non-ionizing photons remains
proportional to age, because of accumulation effects of old stars
which are the major contributors to this spectral range. {\em
(Right)\/}: Evolution of an impulsive star formation burst
(lasting~$10^5$\,yr), for the same extreme stellar compositions as
above. Note that the ionizing photons practically disappear at an
age of $\sim 10$~Myr. Figure from Sternberg et\,al.\ 2002.}\\
% ----------------------------------------------------------------

The evolution of the photon emission rates for photons above
the Lyman ($Q_{\rm H}$) and the \HeI\ ($Q_{\rm He}$) ionization
thresholds is shown in Figure~11 for both modes of the clusters.
The key difference between the two modes leads to completely
different shapes of the photon emission rates, which will be easy to
distinguish in view of their photoionizing properties acting on their
gaseous environments. It is also shown that increasing the upper IMF
mass cut-off increases the photon emission rates, but does not change
the shapes of the rates obtained during the evolution. The methods
developed for determining extragalactic abundances and population
histories from an analysis of \HII~regions are thus very promising.

From these investigations it is obvious that realistic spectral
energy distributions of massive stars and stellar clusters are
important for studies of their environments. The crucial question,
however, whether the spectral energy distributions of
massive\linebreak
% ----------------------------------------------------------------
\myfigure{\includegraphics[angle=90,width=10.5cm]{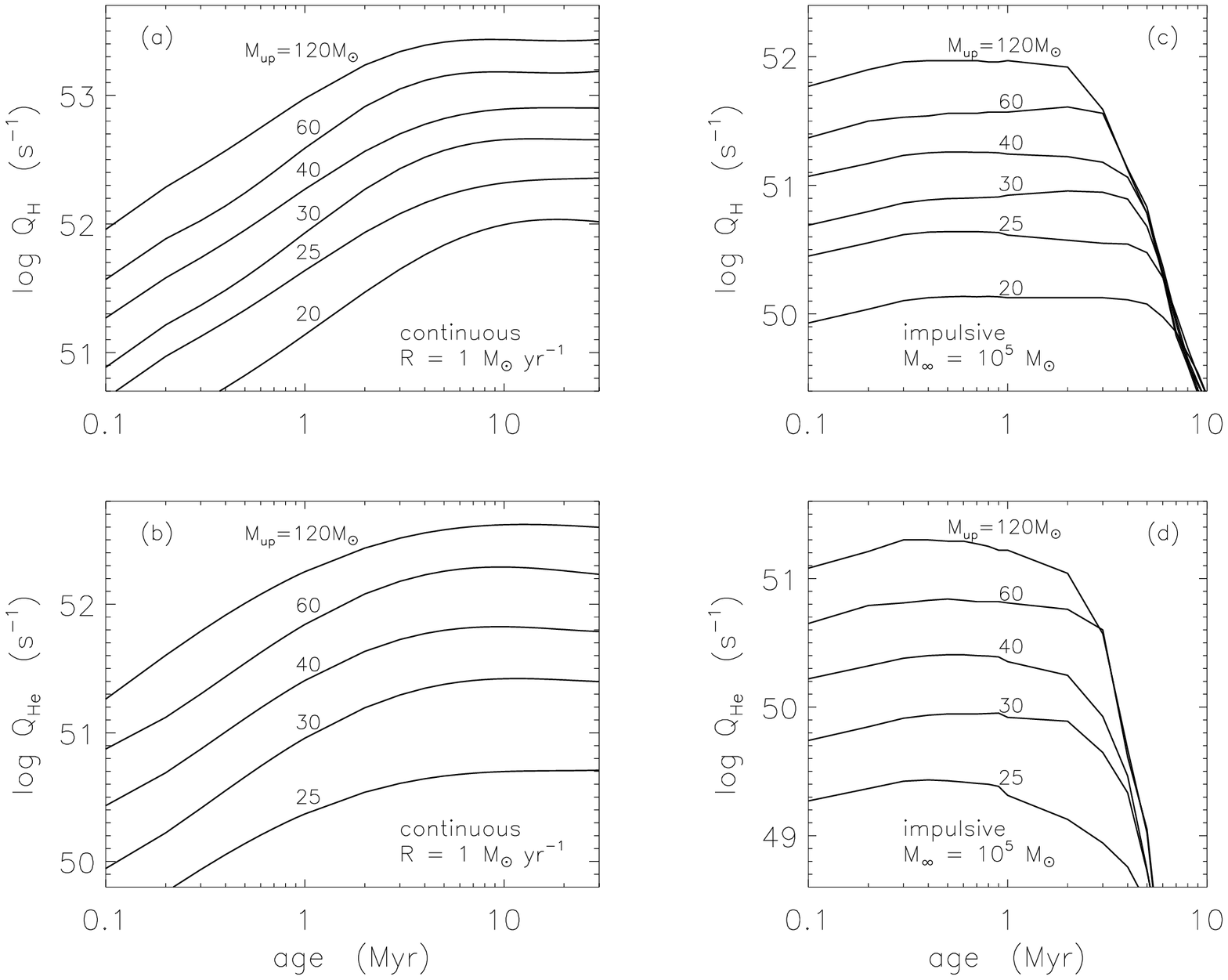}}
Figure 11: {\small Evolution of the Lyman and \HeI\ photon emission
rates for a range of values of the upper IMF mass cut-off. The
continuous star formation mode is considered on the left hand side,
and the impulsive star formation burst on the right. Figure from
Sternberg et\,al.~2002.}\\
% ----------------------------------------------------------------

\noindent stars are already realistic enough to be used for
diagnostic issues of \HII~regions has not been answered yet, since
on the basis of the results obtained for the diagnostic diagram in
Section~4.1 it cannot be excluded that wrong fluxes could show the
same improvements in the Ne$^{++}$/Ne$^+$ ratios just by chance.

\medskip \noindent
An answer to this question requires an {\it ultimate test\/}!

\noindent
This ultimate test is only provided by comparing the observed
and synthetic UV spectra of the individual massive stars, based
on the following reasons:

\smallskip \noindent
1. This test involves hundreds of spectral signatures of various
ionization stages with different ionization thresholds which cover
a large frequency range.

\smallskip \noindent
2. Almost all of the ionization thresholds lie within the spectral
range shortward of the Lyman ionization threshold (cf.~Figure~12);
thus, {\em the ionization balances of all elements depend
sensitively on the ionizing radiation throughout the entire wind.}

\medskip
The ionization balance can be traced reliably through the strength
and structure of the wind lines formed throughout the atmosphere.
Hence, it is a natural and the only reliable step to test the
quality of the ionizing fluxes by virtue of their direct product:
{\it the UV~spectra of O~stars}.

But before we turn to this test we will continue our discussion
with another astrophysically important stellar object.

% ----------------------------------------------------------------
\myfigure{\includegraphics[width=11.0cm]{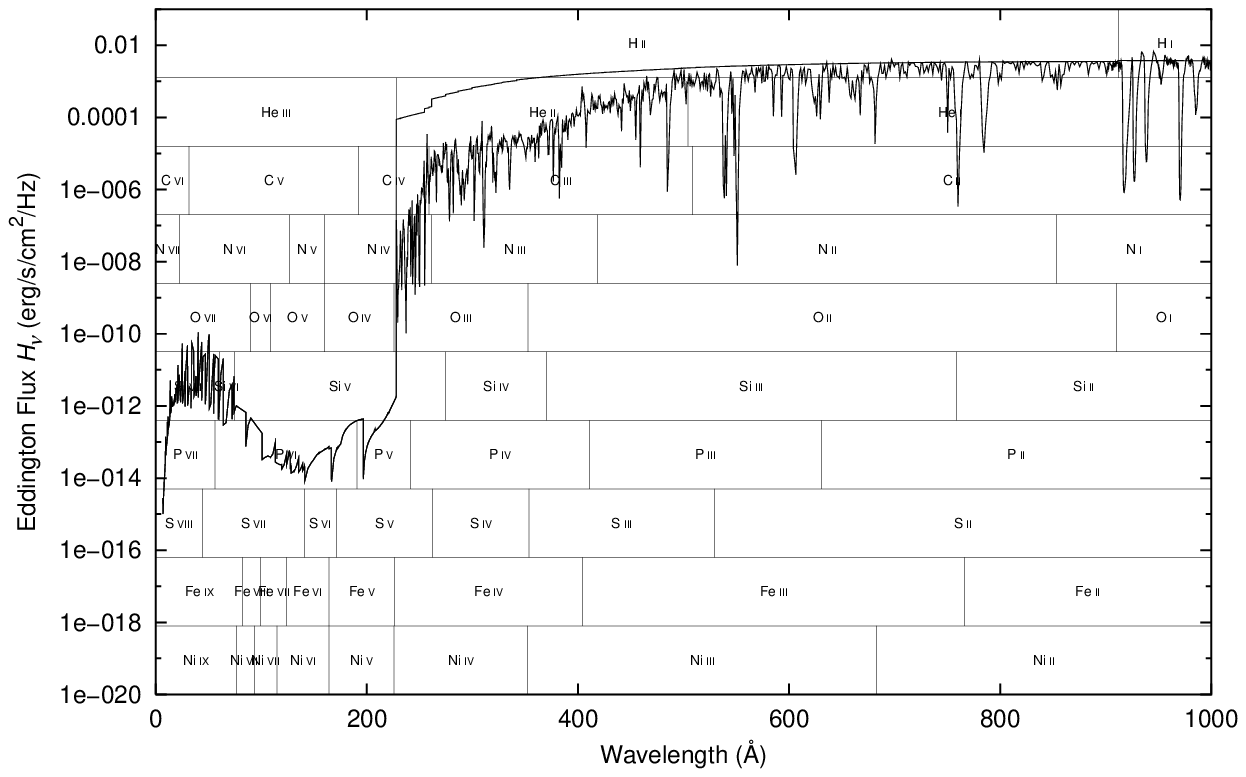}}
Figure 12: {\small Spectral energy distribution shortward of the
Lyman ionization threshold of an expanding NLTE model atmosphere
for an O~supergiant (Pauldrach et\,al.\ 2003). The small vertical
bars indicate the ionization thresholds for all important ions; the
ionization balance depends almost entirely on the ionizing flux,
and this influence can be traced by the spectral lines in the
observable part of the UV~spectrum.}\\
% ----------------------------------------------------------------

\section{Supernovae of Type~Ia as Distance Indicators}

In order to complete the discussion of Hot Stars in the context of
modern astronomy, we will concentrate now on the
subject of Supernovae of Type~Ia. With respect to diagnostic issues
we will investigate the role of Supernovae of Type~Ia as distance
indicators. The context of this discussion regards, as a starting
point, the current surprising result that distant SNe~Ia at
intermediate redshift appear fainter than standard candles in an
empty Friedmann model of the universe.

Type~Ia supernovae, which are the result of the thermonuclear
explosion of a compact low mass star, are currently the best known
distance indicators. Due to their large luminosities ($L_{\rm
max} \sim 10^{43}$\,erg/s) they reach far beyond the local supercluster
(cf.~Saha et\,al.\ 2001 and references therein). It is thus not
surprising that Type~Ia supernovae have become the most important
cosmological distance indicator over the last years, and this is
not only due to their extreme brightness, but primarily due to
their maximum luminosity which can be normalized by their light
curve shape, so that these objects can be regarded as {\it standard
candles}. This has been shown by
observations of local SNe~Ia which define the linear expansion of
the local universe extremely well (Riess et\,al.\ 1999), which
in turn is convincing proof of the accuracy of the measured distances. On
the other hand, observations of distant SNe~Ia (up to redshifts of
about 1) have yielded strong evidence that the expansion rate has
been accelerated 6~Gyr ago (cf.~Riess et\,al.\ 1998 and Perlmutter
et\,al.\ 1999).

% ----------------------------------------------------------------
\myfigure{\includegraphics[height=7.2cm]{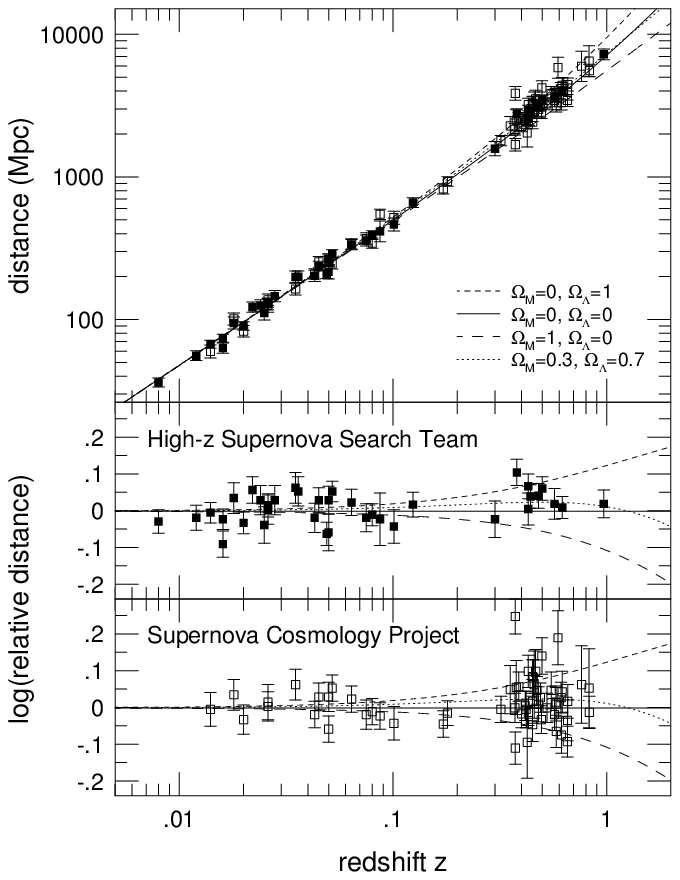}\hfill
\includegraphics[height=7.2cm]{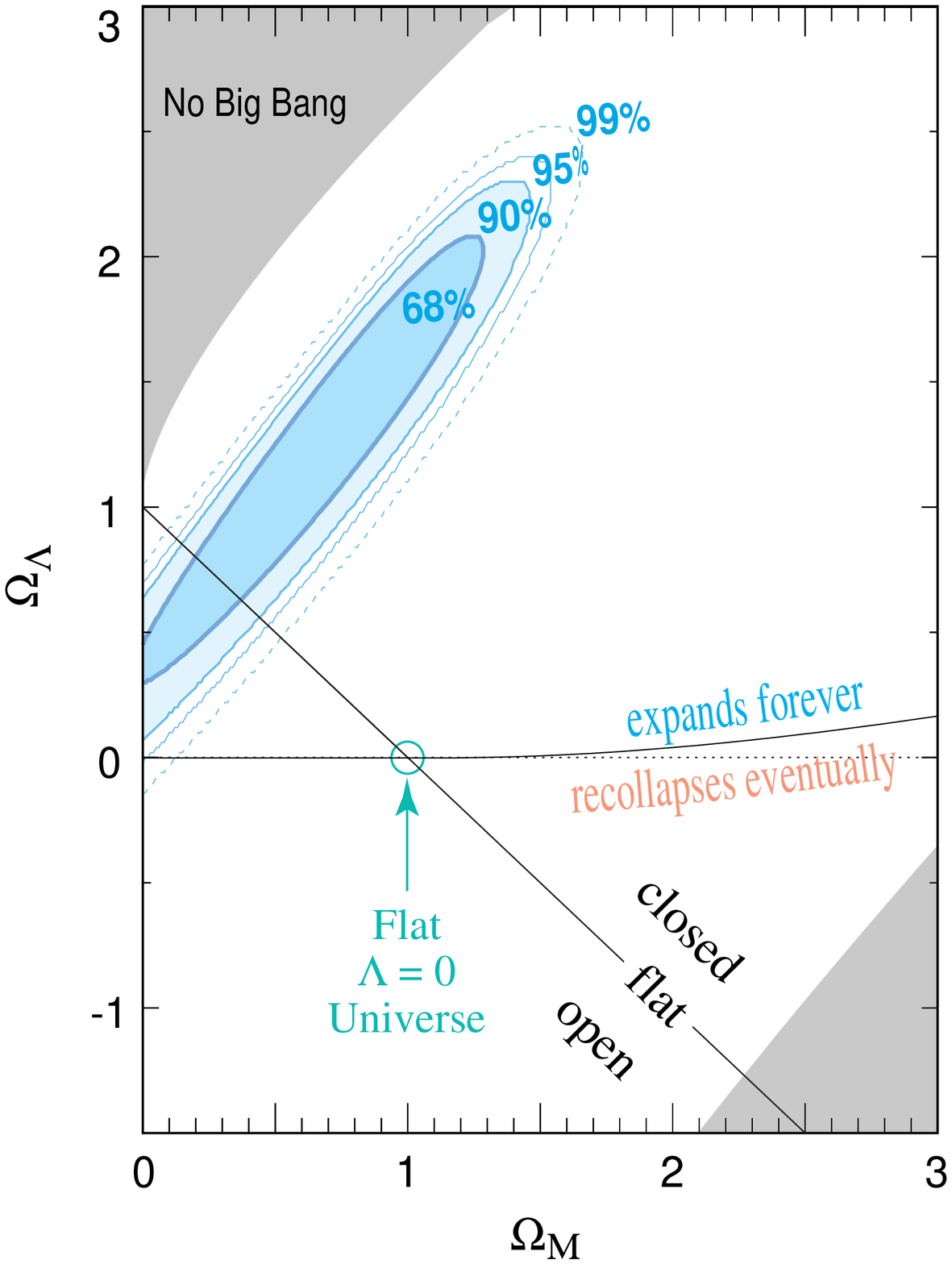}}
Figure 13: {\small {\em Left\/}:~Hubble diagram of Type~Ia
Supernovae. In the upper panel the distance modulus versus redshift
is shown in the usual way, whereas the distance modulus has been
normalized to an empty universe in the lower panels. Shown are the
data of the High-$z$ SN~Search Team (filled squares; Riess et\,al.\
1998) and the Supernova Cosmology Project (open squares; Perlmutter
et\,al.\ 1999). In addition, lines of four cosmological models are
drawn, where the full line corresponds to a cosmological model for
an empty universe and the dotted line to a flat universe. Figure
from Leibundgut~2001.
\newline
{\em Right\/}:~Cosmological diagram showing the likelihood region
as defined by SNe~Ia in the $\Omega_{\Lambda}$ versus $\Omega_{\rm
Mass}$ plane. The results of 79~SNe~Ia have been included -- 27
local and 52 distant ones -- from the sources above. Contours
indicate 68\,\%, 90\,\%, 95\,\%, and 99\,\% probability, and the
line for a flat universe is indicated. Note that the
SN~Ia~luminosity distances clearly indicate an accelerated
expansion of the universe. Figure from Perlmutter
et\,al.~1999.}\\
% ----------------------------------------------------------------

This surprising result becomes evident if we look at the
Hubble diagram in the form of distance modulus versus redshift
(cf.~Figure~13). As is shown in the diagrams of the lower panels,
which are normalized to a cosmological model for an empty universe,
most SNe~Ia at intermediate redshift are positioned at positive
values of the normalized distance modulus; thus, the distant
supernovae appear fainter than what would be expected in a empty
universe. This means that deceleration from gravitational action
of the matter content does not take place.  Moreover, the SNe Ia
appear even more distant indicating an accelerated expansion over
the last 6~Gyr.

The implications of this are immediately recognized in the
cosmological diagram on the right-hand side of Figure~13, where
$\Omega_{\Lambda}$ versus $\Omega_{\rm Mass}$ is plotted. The SN~Ia
results obviously exclude any world model with matter but without a
a cosmological constant. Hence, the data show the need for an
energy contribution by the vacuum. Thus, the SN~luminosity
distances indicate accelerated expansion of the universe!

However, this is not the only interpretation of the result
obtained. There are other astrophysical explanations, such as
obscuration by intergalactic dust or evolution of\linebreak
% ----------------------------------------------------------------
\myfigure{\includegraphics[width=10.cm]{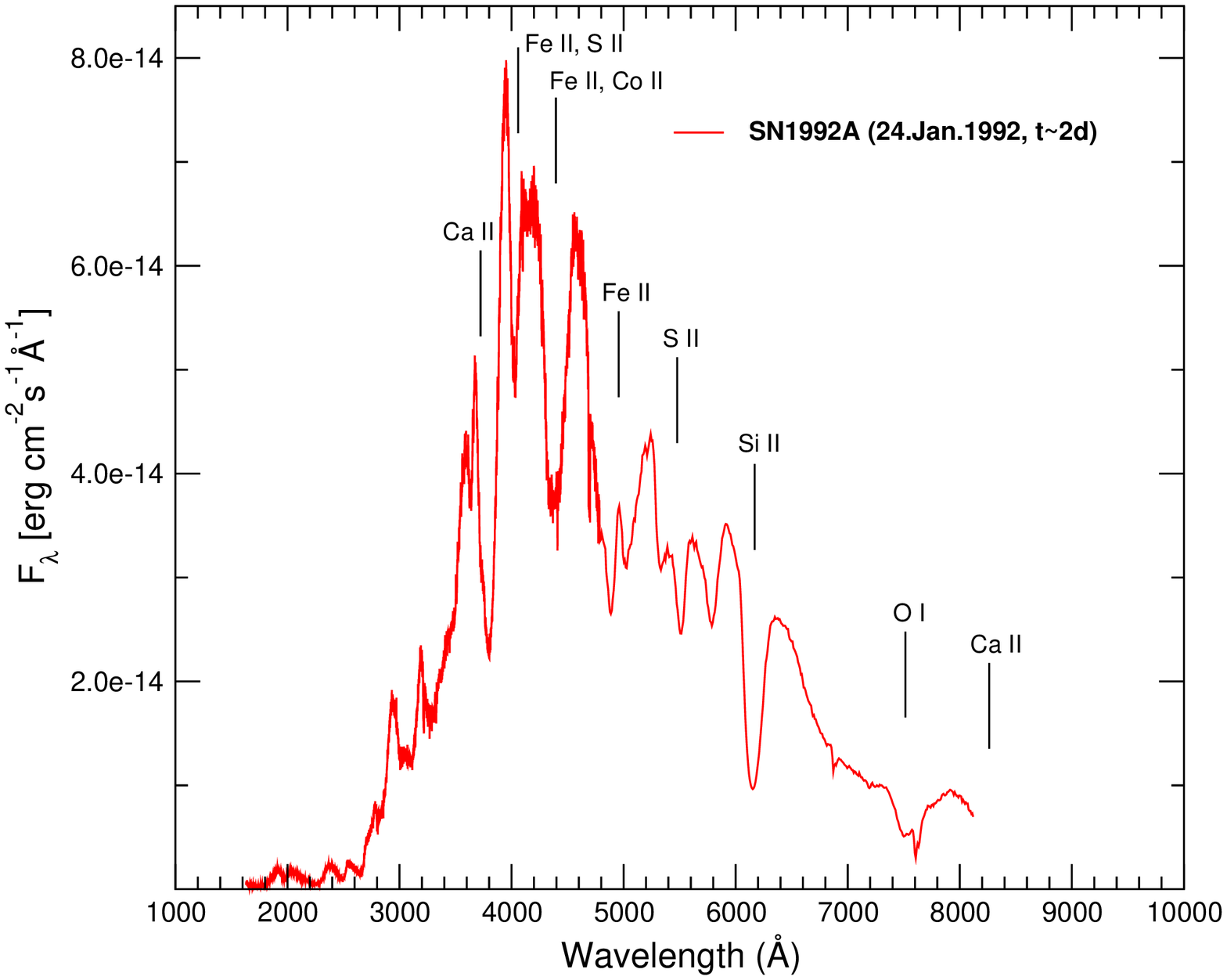}}
Figure 14: {\small Observed HST spectrum of a standard Supernova of
Type~Ia at early phase -- SN~1992A from 24~Jan.~1992. Figure from
Kirshner et\,al.~1993.}\\
% ----------------------------------------------------------------

\noindent supernovae of Type~Ia (cf.~Leibundgut~2001). The latter
point refers primarily to the absolute peak luminosity of SNe~Ia
which might have changed between the local epoch and a redshift
of 1.0, especially in the case of a decrease: then the fainter
supernovae would not be a signature of larger dis\-tances, but rather
of an evolution of these objects.  Although present SN~Ia models do
not favor such an evolution, it cannot be excluded, because neither
the explosions nor the radiation emerging from the atmosphere are
understood in detail yet. Consequently, serious caveats for the
cosmological interpretation of distant supernovae exist.

Thus, we are faced with the question: Are SNe~Ia standard candles
independent of age, or is there some evolution of the SN
luminosity with age? Among other things spectroscopy is certainly a
powerful tool to obtain an answer to this question by searching for
spectral differences between local and distant SNe~Ia.

Unfortunately, at present, we do not have a clear picture of the
exact physical processes which take place in a SN~Ia explosion and
how the radiation released from it should be treated
(for a recent review see Hillebrandt and Niemeyer~2000). In
order to get an impression of the basic physics involved that
affect the atmospheric structure of a SN~Ia, a quick glance at an
observed spectrum will help to point out the necessary
steps for their further analysis.

Figure~14 shows a typical SN~Ia spectrum at early epochs ($< 2$
weeks after maximum). The most striking features of this spectrum
are the characteristic P~Cygni line profiles which are quite similar
to the signatures of O~stars. But compared to the latter objects
there are also important differences: the broad lines indicate that
the velocities of the SN~Ia ejecta are almost a factor of 10 larger
(up to 30\,000~km/s) and, as a second important point, SN~Ia spectra
contain no H and He lines.  Instead, prominent absorption features
of mainly intermediate-mass elements (\SiII, \OI, \SII, \CaII, \MgII,
\dots) embedded in a non-thermal pseudo-continuum are observed.

To answer the question of whether SN~Ia are standard candles in a
cosmological sense, realistic models and synthetic spectra of
Type~Ia Supernovae are obviously required for the analysis of the
observed spectra. As just shown, however, these models will
necessarily need to be based on a similar serious approach for
expanding atmospheres, characterized by non-equilibrium
thermodynamics, as is the case for Hot Stars in general.

It will be shown in Section~7.1 that such spectra are in principle
already available.

\section{Concept for Consistent Models\\ of Hot Star Atmospheres}

In order to determine stellar abundances, parameters, and physical
properties (and from these, obtain realistic spectral energy
distributions) of Hot Stars via quantitative UV~spectroscopy, a
principal difficulty needs to be overcome: {\it the diagnostic
tools and techniques must be provided}. This requires the
construction of detailed atmospheric models.

As has been demonstrated in the previous sections, such a tool has
not been made obsolete by general development in astrophysics. On the
contrary, it is becoming more and more relevant to current
astronomical research as spectral analysis of hot luminous stars is
of growing astrophysical interest. Thus, continuing effort is
expended to develop a standard code in order to provide the necessary
diagnostic tool. In the following part of this section we describe
the status of our continuing work to construct realistic models for
expanding atmospheres.

Before we focus on the theory in its present stage, we should
mention previous fundamental work which turned out to be essential
in elaborating the theory. In this context I want to emphasize
publications which refer to key aspects of theoretical activity. The
starting point of the development of the radiation-driven wind
theory is rooted in a paper by Milne~(1926) more than 70 years ago.
Milne was the first to realize that radiation could be coupled to
ions and that this process subsequently may eject the ions from the
stellar surface. The next fundamental step goes back to
Sobolev~(1957), who developed the basic ideas of radiative transfer
in expanding atmospheres.

Radiation pressure as a driving mechanism for stellar outflow was
rediscovered by Lucy and Solomon~(1970) who developed the basis of
the theory and the first attempt to its solution. The pioneering step
in the formulation of the theory in a quasi self-consistent manner
was performed by Castor, Abbott, and Klein~(1975). Although these
approaches were only qualitative (due to many simplifications),
the theory was developed further owing to the promising results
obtained by these authors.  Regarding key aspects of the solution
of the radiative transfer the work of Rybicki~(1971) and Hummer and
Rybicki~(1985) has to be emphasized. With respect to the soft X-ray
emission of O~stars (detected by Seward et\,al.\ 1979 and Harnden
et\,al.\ 1979), Cassinelli and Olson~(1979) have investigated the
possible influence of X-rays\linebreak
% ----------------------------------------------------------------
\myfigure{\includegraphics[width=\textwidth]{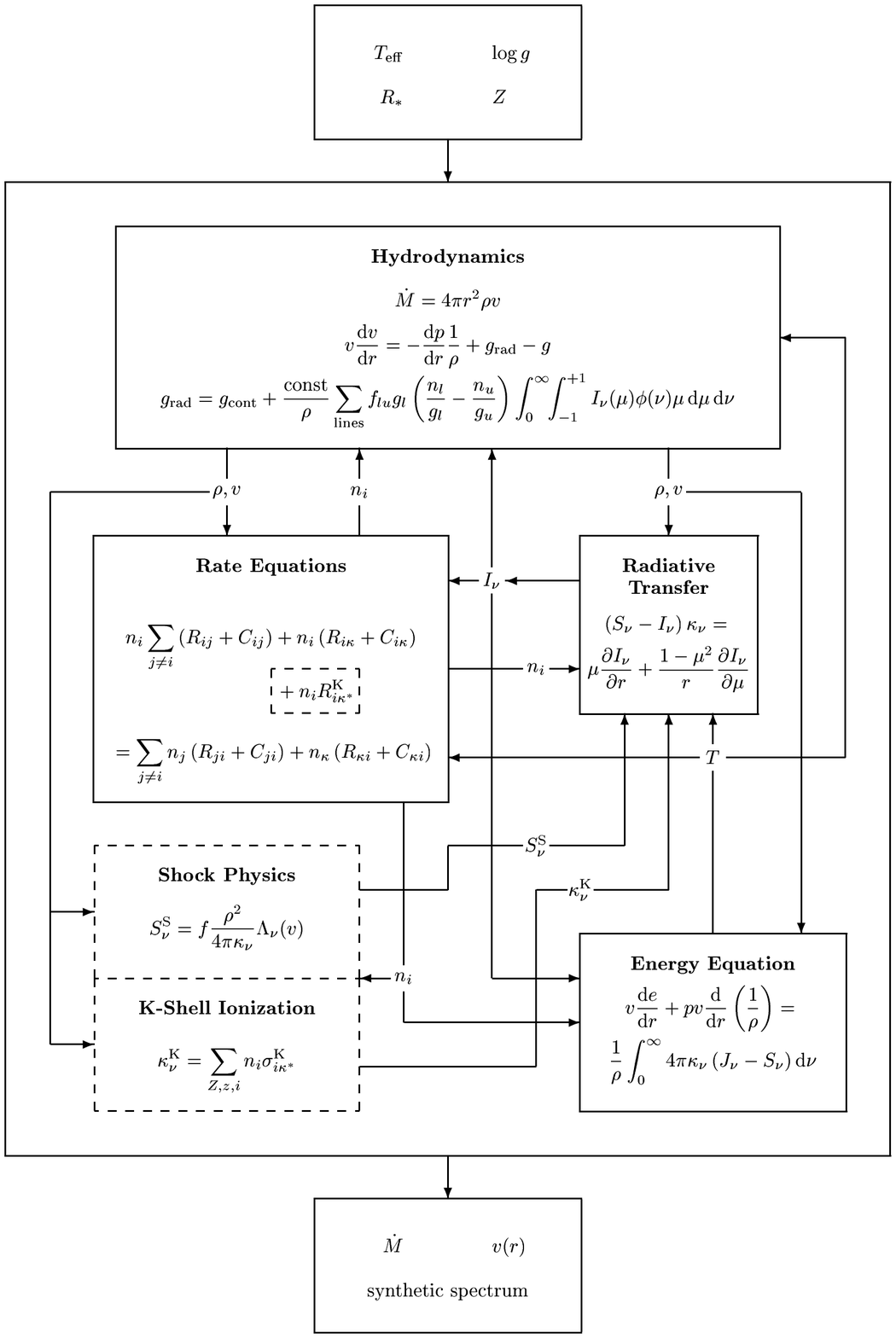}}
Figure 15: {\small Schematic sketch of the basic
equations and the non-linear system of integro-differential
equations that form the basis of stationary models of hot star
atmospheres -- for details see text. Figure from Pauldrach
et\,al.\ 1994b).}
% ----------------------------------------------------------------

\noindent on the ionization structure. Finally,
Lucy and White~(1980) and Owocki, Castor, and Rybicki~(1988) have
distinguished themselves with basic theoretical investigations of
time-dependent radiation hydrodynamics which describe the creation
and development of shocks (for more details about the role of X-rays
in the atmospheres of Hot Stars see the reviews of Pauldrach et\,al.\
1994b and Kudritzki and Puls~2000).

\subsection{The general method}

The basis of our approach in constructing detailed atmospheric
models for Hot Stars is the concept of {\em homogeneous,
stationary, and spherically symmetric radiation driven winds},
where the expansion of the atmosphere is due to scattering and
absorption of Doppler-shifted metal lines. Although these
approximations seem to be quite restrictive, it has already been
shown that the time-averaged mean of the observed UV spectral
features can be described correctly by such a method (Pauldrach
et\,al.\ 1994, 1994b).

The primary output of this kind of model calculation are the
spectral energy distributions and synthetic spectra emitted by the
atmospheres of hot stars. As these spectra consist of hundreds of
strong and also weak wind-contaminated spectral lines
which form the basis of a quantitative analysis, and as the
spectral energy distribution from hot stars is also used as input
for the analysis of emission line spectra which depend sensitively
on the structure of the emergent stellar flux (cf.~Section~4), a
sophisticated and well-tested method is required to produce these
data sets accurately.

However, developing such a method is not straightforward, since
modelling the atmospheres of hot star involves the replication
of a tightly interwoven mesh of physical processes: the equations
of radiation hydrodynamics including the energy equation, the statistical
equilibrium for all important ions with detailed atomic physics,
and the radiative transfer equation at all transition frequencies
have to be solved simultaneously. Figure~15 gives an overview of
the physics to be treated.

\medskip
\noindent
The principal features are:

\medskip
\arabiclist
\item {\em The stellar parameters\/} $\Teff$ (effective
temperature), $\logg$ (logarithm of photospheric gravitational
acceleration), $R_*$ (photospheric radius defined at a Rosseland
optical depth of 2/3) and $Z$ (abundances) have to be
pre-specified.
\item {\em The hydrodynamic equations\/} are solved ($r$
is the radial coordinate, $\varrho$ the mass density, $v$ the
velocity, $p$ the gas pressure and $\Mdot$ the mass loss rate). The
crucial term is the radiative acceleration $g_{\rm rad}$ with
minor contributions from continuous absorption and major
contributions from scattering and line absorption. For each line
the oscillator strengths $f_{lu}$, the statistical weights $g_l$,
$g_u$, and the occupation numbers $n_l$, $n_u$ of the lower and
upper level enter the equations, together with the frequency and angle integral
over the specific intensity $I_\nu$ and the line broadening
function $\varphi_\nu$ accounting for the Doppler effect.
\item {\em The occupation numbers\/} are determined by the
{\em rate equations\/} containing collisional ($C_{ij}$) and
radiative ($R_{ij}$) transition rates. Low-temperature dielectronic
recombination is included and Auger ionization due to K-shell
absorption (essential for C, N, O, Ne, Mg, Si, and S) of soft X-ray
radiation arising from shock-heated matter is taken into account
using detailed atomic models for all important ions. Note that the
hydrodynamical equations are coupled directly with the rate
equations. The velocity field enters into the radiative rates while
the density is important for the collisional rates and the equation
of particle conservation. On the other hand, the occupation numbers
are crucial for the hydrodynamics since the radiative line
acceleration dominates the equation of motion.
\item {\em The spherical transfer equation\/} which yields
the radiation field at every depth point, including the thermalized
layers where the diffusion approximation is applied, is solved for
the total opacities ($\kappa_\nu$) and source functions ($S_\nu$)
of all important ions. Hence, the influence of the spectral lines
-- the strong EUV {\em line blocking\/} -- which affects the
ionizing flux that determines the ionization and excitation of
levels, is taken into account. Note that the radiation field is
coupled with the hydrodynamics ($g_{\rm rad}$) and the rate
equations ($R_{ij}$).

Moreover, the {\em shock source functions\/} ($S_\nu^{\rm S}$)
produced by radiative cooling zones which originate from a
simulation of shock heated matter, together with K-shell absorption
($\kappa_\nu^{\rm K}$), are also included in the radiative
transfer. The shock source function is incorporated on the basis of
an approximate calculation of the volume emission coefficient
($\Lambda_\nu$) of the X-ray plasma in dependence of the
velocity-dependent post-shock temperatures and the filling factor
($f$).
\item {\em The temperature structure\/} is determined by the
microscopic {\em energy equation\/} which, in principle, states
that the luminosity must be conserved. {\em Line blanketing\/}
effects which reflect the influence of line blocking on the
temperature structure are taken into account.
\item  The iterative solution of the total system of equations
yields the hydrodynamic structure of the wind (i.\,e., the {\em
mass-loss rate\/} $\Mdot$ and the {\em velocity structure\/}
$v(r)$) together with {\em synthetic spectra\/} and ionizing
fluxes.
\end{list}

\medskip

\noindent
Essential steps which form the basis of the theoretical framework
developed are described in Pauldrach et\,al.\ 1986, Pauldrach~1987,
Pauldrach and Herrero~1988, Puls and Pauldrach~1990, Pauldrach
et\,al.\ 1994, Feldmeier et\,al.\ 1997, and Pauldrach et\,al.\
1998.

The effect complicating the system the most is the overlap of the
spectral lines. This effect is induced by the velocity field of the
expanding atmosphere which shifts at different radii up to 1000
spectral lines of different ions into the line of sight at each
observer's frequency. Since the behavior of most of the UV spectral
lines additionally depends critically on a detailed and consistent
description of the corresponding effects of {\em line blocking\/}
and {\em line blanketing\/} (cf.~Pauldrach~1987), special attention
has to be given to the correct treatment of the Doppler-shifted
line radiation transport of all metal lines in the entire sub- and
supersonically expanding atmosphere, the corresponding coupling
with the radiative rates in the rate equations, and the energy
conservation. Another important point to emphasize concerns the
atomic data, since it is quite obvious that the quality of the
calculated spectra, the multilevel NLTE treatment of the metal ions
(from~C to~Zn), and the adequate representation of line blocking
and the radiative line acceleration depends crucially on the
quality of the atomic models. Thus, the data have to be continuously
improved whenever significant progress of atomic data modelling
is made. Together with a revised inclusion of EUV and X-ray
radiation produced by cooling zones which originate from the
simulation of shock heated matter, these improvements have been
implemented and described in a recent paper (Pauldrach et\,al.\
2001). Very recently we have additionally focused on a remaining
aspect regarding hydrodynamical calculations which are consistent
with the radiation field obtained from the line-overlap
computations, and we have incorporated this improvement into
our procedure (see Pauldrach and Hoffmann~2003).

This solution method is in its present stage already regarded as a
standard procedure towards a realistic description of stationary
wind models. Thus, together with an easy-to-use interface and an
installation wizard, the program package {\em WM-basic\/} has been
made available to the community. The package can be downloaded
from the URL given on the first page.

In the next section we will investigate thoroughly
whether these kind of models already lead to realistic results.

\section{Synthetic Spectra and Models\\ of Hot Star Atmospheres}

The first five sections of this review inevitably lead to the
realization that line features of expanding atmospheres are one of
the most important astronomical manifestations, since they can
easily be identified even in spectra of medium resolution in
individual objects like supernovae or in integrated spectra of
starburst regions even at significant redshift. As a consequence,
they can be used to provide important information about the
chemical composition, the extragalactic distance scale, and several
other properties of miscellaneous stellar populations. All that is
required are {\it hydrodynamic NLTE model atmospheres\/} that
incorporate the effects of spherical extension and expansion as
realistic and consistent as possible. After almost two decades of
work these model atmospheres now seem to be available, and in the
following we will investigate thoroughly whether basic steps
towards realistic synthetic spectra for Supernovae of Type~Ia can
already be presented, and whether UV~diagnostic techniques applied
to two of the most luminous O~supergiants in the Galaxy lead to
synthetic spectra which can be regarded as a measure of excellent
quality.

\subsection{Synthetic and Observed Spectra\\ of Supernovae of Type~Ia}

Prominent features in the spectra of Type~Ia supernovae in early
phases are the characteristic absorption lines resulting from low
ionized intermediate-mass elements, such as \SiII, \OI, \CaII,
\MgII, cf.~Figure~17. These absorption lines show characteristic
line shapes due to Doppler-broadening resulting from the large
velocity gradients. The formation of these lines results in a
pseudo-continuum that is set up\linebreak
% ----------------------------------------------------------------
\myfigure{\includegraphics[width=0.50\textwidth]{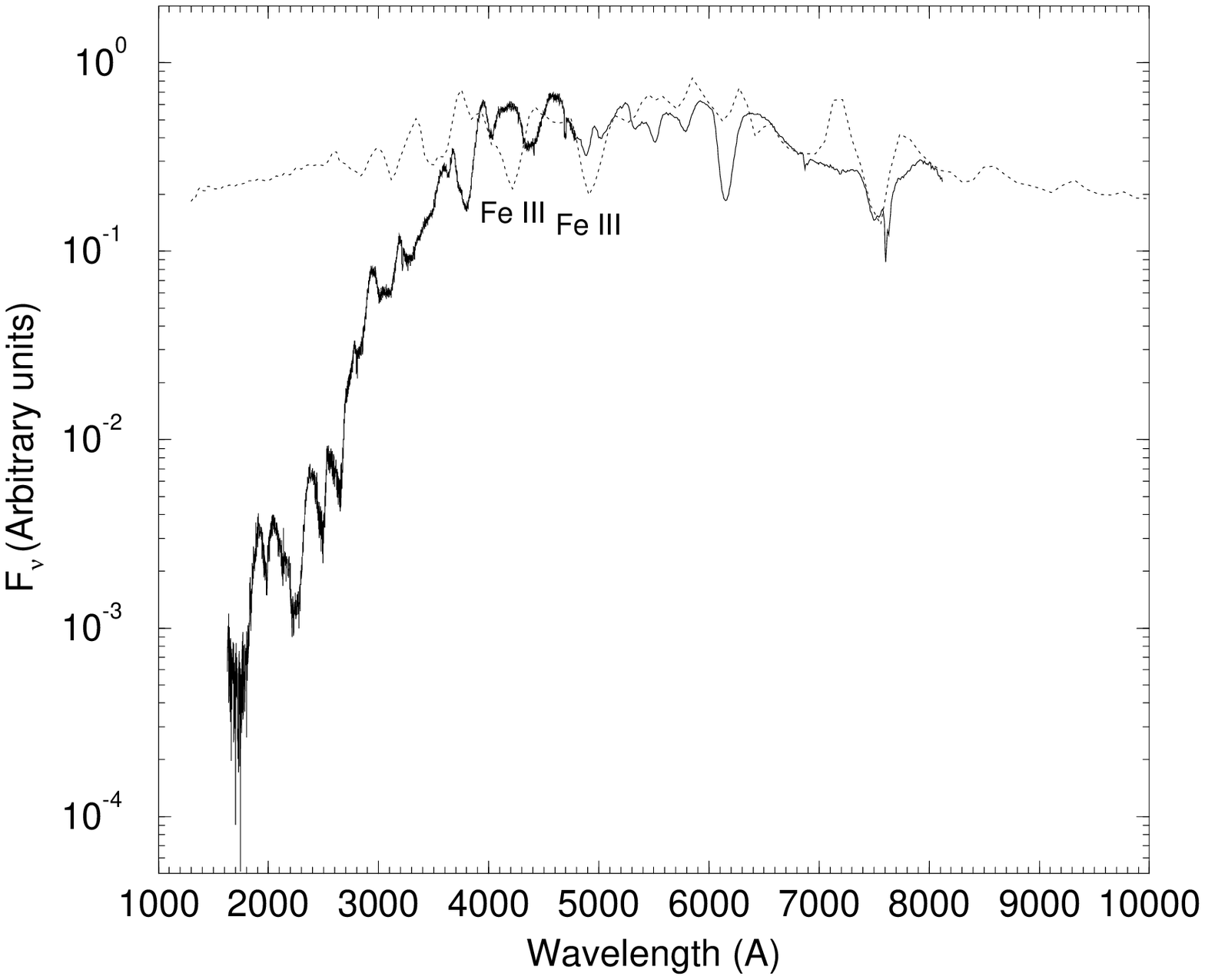}\hfill
\includegraphics[width=0.50\textwidth]{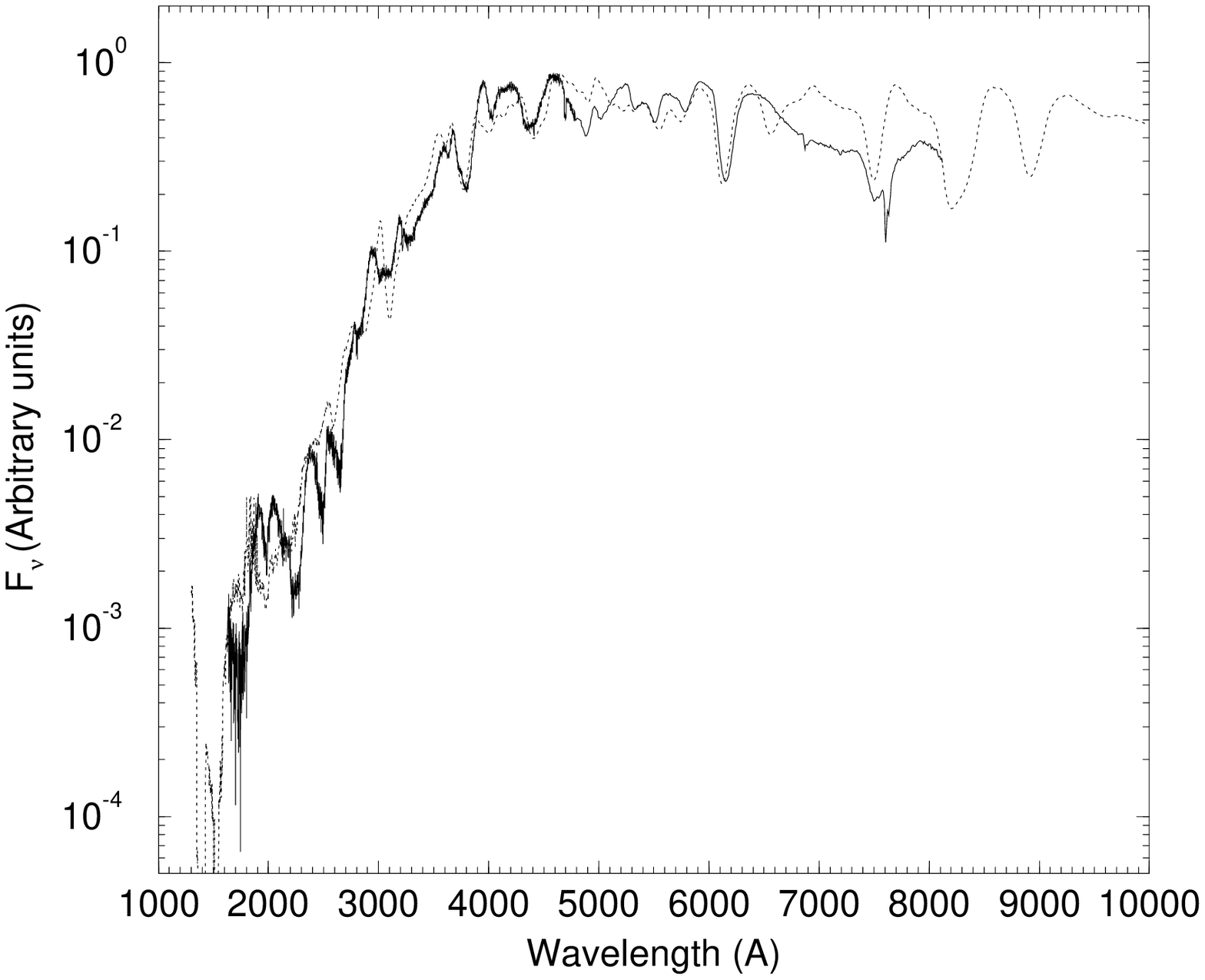}}
\includegraphics[width=5.5cm]{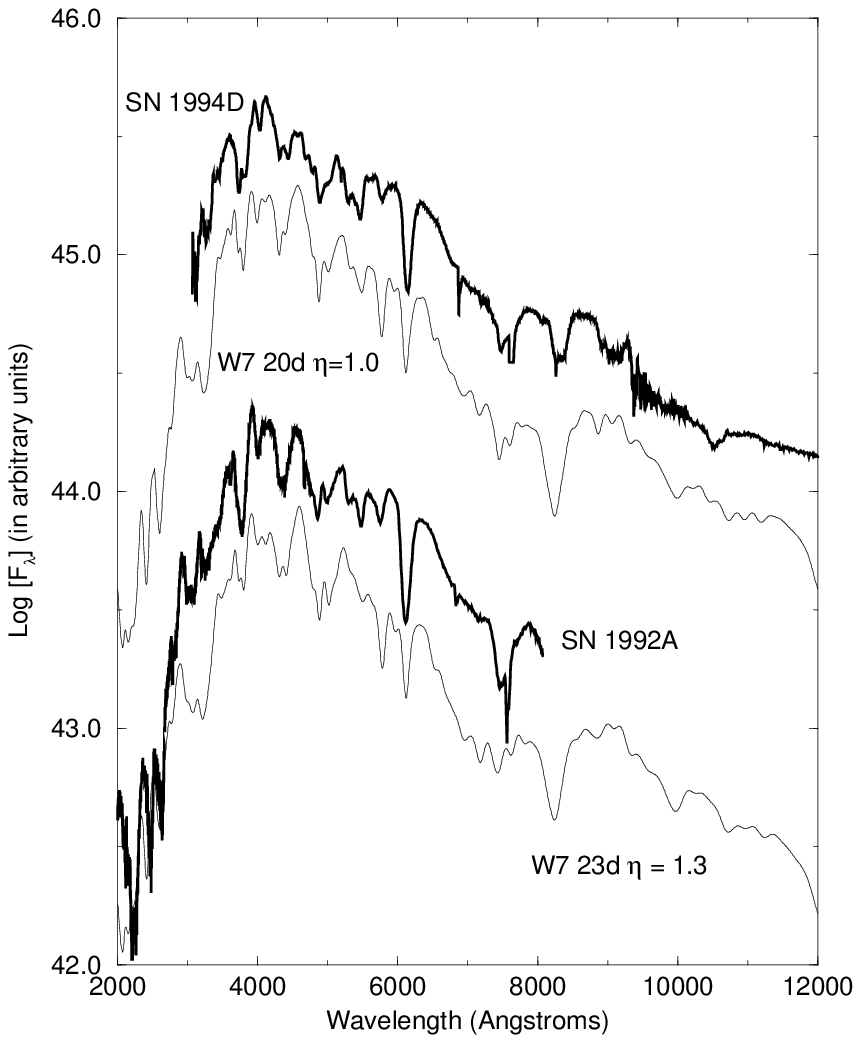}\hfill
\parbox[b]{6.cm}{Figure 16: {\small Synthetic NLTE spectra of
a Supernova of Type~Ia. {\em Top\/}:~The dotted lines show the
spectrum from a model without line blocking (left panel) and from
one where line blocking is treated consistently (right panel), both
compared to the observed spectrum of SN~1992A (solid line). (Figure
from Pauldrach et\,al.\ 1996.) {\em Left\/}:~The two curves in the
lower part show a synthetic spectrum for a W7 model at 23 days
compared to the observed spectrum of SN~1992A. (Figure from Nugent
et\,al.\ 1997.) The observed spectrum is from Kirshner et\,al.\
1993 (5~days after maximum light).}}\\
% ----------------------------------------------------------------

\noindent by the enormous number of these lines. Supernovae
appear in this `photospheric' epoch for about one month after
the explosion and the spectra at this epoch contain useful
information on the energetics of the explosion (luminosity,
velocity- and density-structure) and on the nucleosynthesis in the
intermediate and outer part of the progenitor star. Therefore, the
primary objective in order to analyze the spectra is to construct
consistent models which link the results of the nucleosynthesis
and hydrodynamics obtained from state-of-the-art explosion models
(cf.~Reinecke et\,al.\ 1999) with the calculations of light curves
and synthetic spectra of SN~Ia. In a first step, however, this is
done by using the standard hydrodynamical model W7 by Nomoto et\,al.\
1984. Figure~16 exemplifies the status quo of the synthetic Spectra
of Type~Ia Supernovae at early phases obtained from this first step,
compared to observations.

Obviously, the first comparison shown (top left panel of Figure~16)
does not at all represent a realistic atmospheric model.  The
synthetic spectrum shown has been calculated for a model without line
blocking, i.\,e., only continuum opacity has been considered in the
NLTE calculation. Thus, the drastic effects of line blocking on the
ionization and excitation and the emergent flux can be verified by
comparing the\linebreak
% ----------------------------------------------------------------
\myfigure{\includegraphics[width=8.2cm]{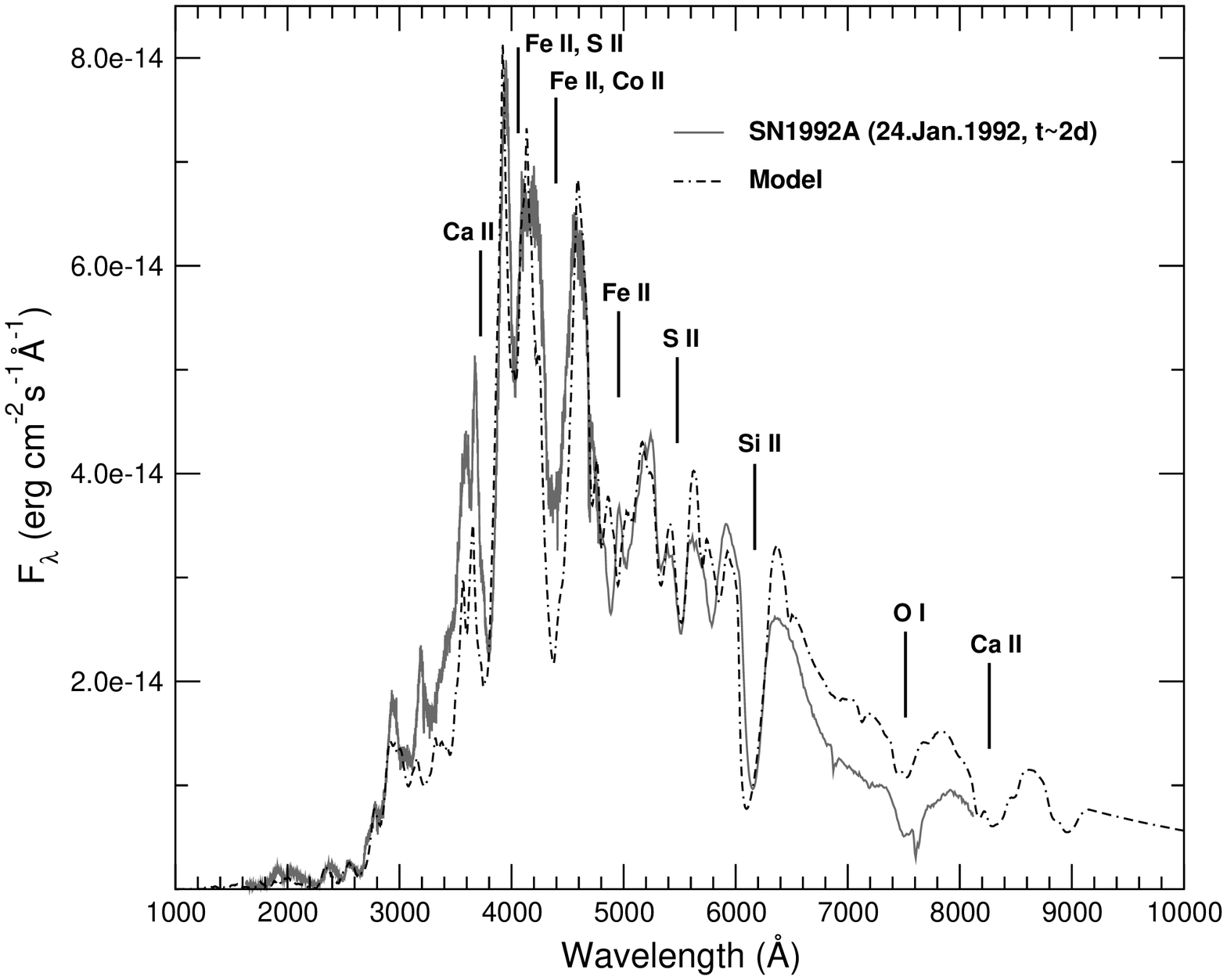}}
Figure 17: {\small Comparison of the most recent synthetic NLTE
spectrum at early phases and the observed spectrum of SN~1992A,
illustrating that the method used is already on a quantitative
level. The rest wavelengths of various metal lines are indicated by
vertical bars. Figure from Sauer and Pauldrach~2002.}\\
% ----------------------------------------------------------------

\noindent calculated flux with the observed flux, where the huge
difference in the UV is due to the missing contribution of the lines
to the opacity. It is an essential by-product of this result that
the observed line features in the optical are also not reproduced:
The two strongest features in the synthetic spectrum are due to
\FeIII\ lines, indicating that the ionization is too high because
of the excessive UV flux. This behavior nicely illustrates that the
ionization balance depends almost entirely on the ionizing UV flux
and that this influence can be traced by the spectral lines in the
observable part of the spectrum, as has been stated in Section~4.2.

The fact that the difference in the UV observed for this model
is indeed due to line blocking becomes obvious if this important
effect is properly taken into account, as in the case of the model
shown in the upper right panel of Figure~16. The agreement of the
resulting spectrum of this model with the observations is quite good,
demonstrating that line blocking alone can diminish the discrepancy
between the synthetic and the observed spectra. Both the flux level
and most line features are now reasonably well reproduced throughout
the spectrum (cf.~Pauldrach et\,al.\ 1996). Additionally, in the
lower left panel of Figure~16 a spectral fit to the same supernova
observation by Nugent et\,al.\ (1997) is shown, which is based on
a completely independent approach to the theory.  The resulting
spectrum is also quite reasonable.

Finally, Figure~17 shows a recently calculated synthetic spectrum
that is based on the important improvements of the theory described
in the previous section, compared to the observed ``standard''
supernova spectrum (SN~1992A).  The synthetic spectrum reproduces
the observed spectrum quite well, and the overall impression of this
comparison is that the method used is already on a quantitative
level, indicating that the basic physics is treated properly
(cf.~Sauer and Pauldrach~2002).

Although there is still need for improvements regarding, for
instance, a proper treatment of the $\gamma$~energy deposition in
the outermost layers and the use of current hydrodynamic explosion
models, these models form the basis for the primary objective of the
subject, i.\,e., to search for spectral differences between local
and distant SNe~Ia. Concerning the facilities of spectral analysis
we are thus close to the point where we can tackle the question of
whether SNe~Ia are standard candles in a cosmological sense.

\subsection{Detailed Analyses of Massive O~Stars}

The objectives of a detailed comparison of synthetic and observed
UV~spectra of massive O~stars are manifold. Primarily it has to be
verified that the higher level of consistent description of the
theoretical concept of line blocking and blanketing effects and
the involved modifications to the models leads to changes in the
line spectra with much better agreement to the observed spectra
than the previous, less elaborated and less consistent models.
Secondly it has to be shown that the stellar parameters, the wind
parameters, and the abundances can be determined diagnostically via
a comparison of observed and calculated high-resolution spectra
covering the observable UV~region. And finally, the quality of
the spectral energy distributions has to be verified.  The latter
point, however, is a direct by-product of the other objectives if
the complete observed high-resolution UV~spectra are accurately
reproduced by the synthetic ones, as illustrated in Section~4.2

In the following, the potential of the improved method will be
demonstrated by an application to the O4\,I(f) star $\zeta$~Puppis
and the O9.5\,Ia star $\alpha$~Cam.

\subsubsection{UV Analysis of the Hot O~Supergiant $\zeta$~Puppis}

Referring to the beginning of the review, we will now start to
examine carefully the old-fashioned UV~spectrum of $\zeta$~Puppis.
To put this into perspective, we will also briefly review the
improvements of UV~line fits following from gradual improvements
of the methods used.

The first serious attempt to analyze the UV~spectrum of this standard
object goes back to Lamers and Morton in~1976. From a present point
of view their model can not be regarded as a sophisticated one,
because they {\em assumed} the complete model structure -- the
dynamical structure, the occupation numbers, and the temperature
structure -- and the line radiative transfer was treated in an
approximative way.  Nevertheless, we learned from this kind of
spectrum synthesis that a solution to the problem is feasible,
and that at least the dynamical parameters -- the mass-loss rate
($\Mdot$) and the terminal velocity ($\vinf$) -- can, in principle,
be determined from UV~lines quite accurately. The UV~line fits
of Hamann in~1980 have been based on the same assumptions apart
from a very detailed treatment of the radiative line transfer.
As a result of this the calculated resonance lines of some ions
were already quite well in agreement with the observed ones. But
the real conclusion to be drawn from these comparisons is that a
detailed analysis requires a more consistent treatment of expanding
atmospheres.

% ----------------------------------------------------------------
\myfigure{\includegraphics[width=9.4cm]{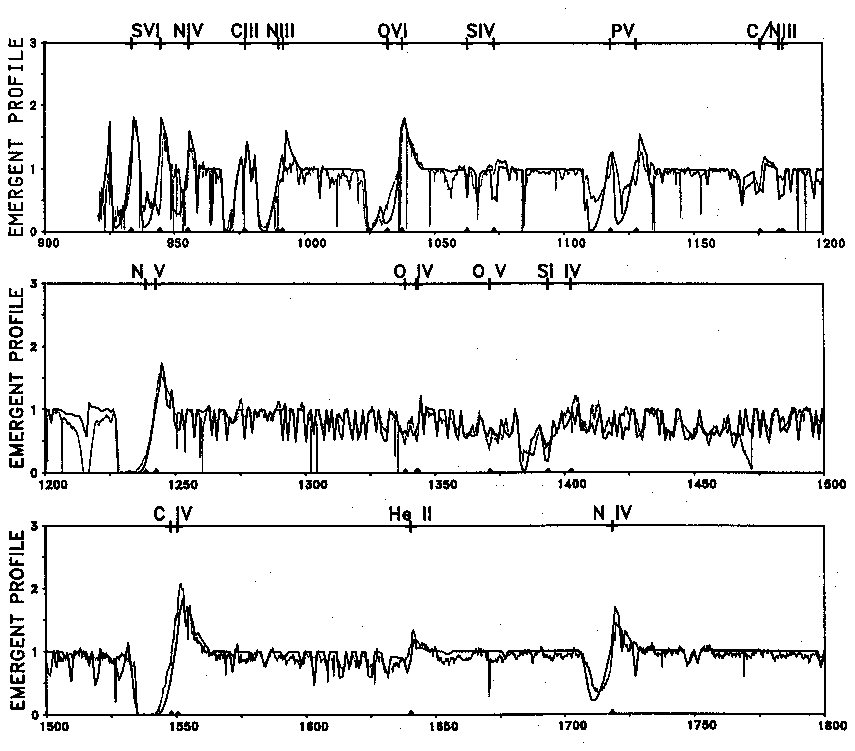}}
Figure 18: {\small Calculated and observed UV spectrum of
$\zeta$~Puppis. The observed spectrum shows the Copernicus and IUE
high-resolution observations, and the calculated spectrum belongs
to the final model of Pauldrach et\,al.\ 1994.}\\
% ----------------------------------------------------------------

The situation improved considerably with the first attempts
to find a consistent solution, meaning that the equations of
hydrodynamics, non-equilibrium thermodynamics, and radiative
transfer have been solved in a consistent way. Pauldrach~(1987)
introduced for the first time a full NLTE treatment of the metal
lines driving the wind, and Puls~(1987) investigated the important
effect of multiple photon momentum transfer through line overlaps
caused by the velocity-induced Doppler-shifts for applications in
stellar wind dynamics. From this procedure, dynamical parameters,
constraints on the stellar parameters, and, as has been verified
by a comparison of a sample of calculated spectral lines with
the observed spectrum, an ionization equilibrium and occupation
numbers which were close to a correct description have already
been obtained. But, among other approximations, the models still
suffered from the neglect of radiation emitted from shock cooling
zones and from a very approximate treatment of line blocking.

The further steps to reproduce most of the important observed
individual line features in the~UV required a lot of effort in
atomic physics, in improved NLTE multilevel radiative transfer,
and in spectrum synthesis techniques. An important step towards this
objective was the paper by Pauldrach et\,al.~(1994), where for the
first time stellar parameters and abundances have been determined
from individual UV line features. Figure~18 shows the corresponding
synthetic spectrum that was already in overall agreement with
the observations.  Nevertheless, the treatment was still affected
by a number of severe approximations regarding especially the
neglect of the line blanketing effect, an approximate treatment
of the line radiative transfer,\linebreak
% ----------------------------------------------------------------
\myfigure{\includegraphics[width=\textwidth]{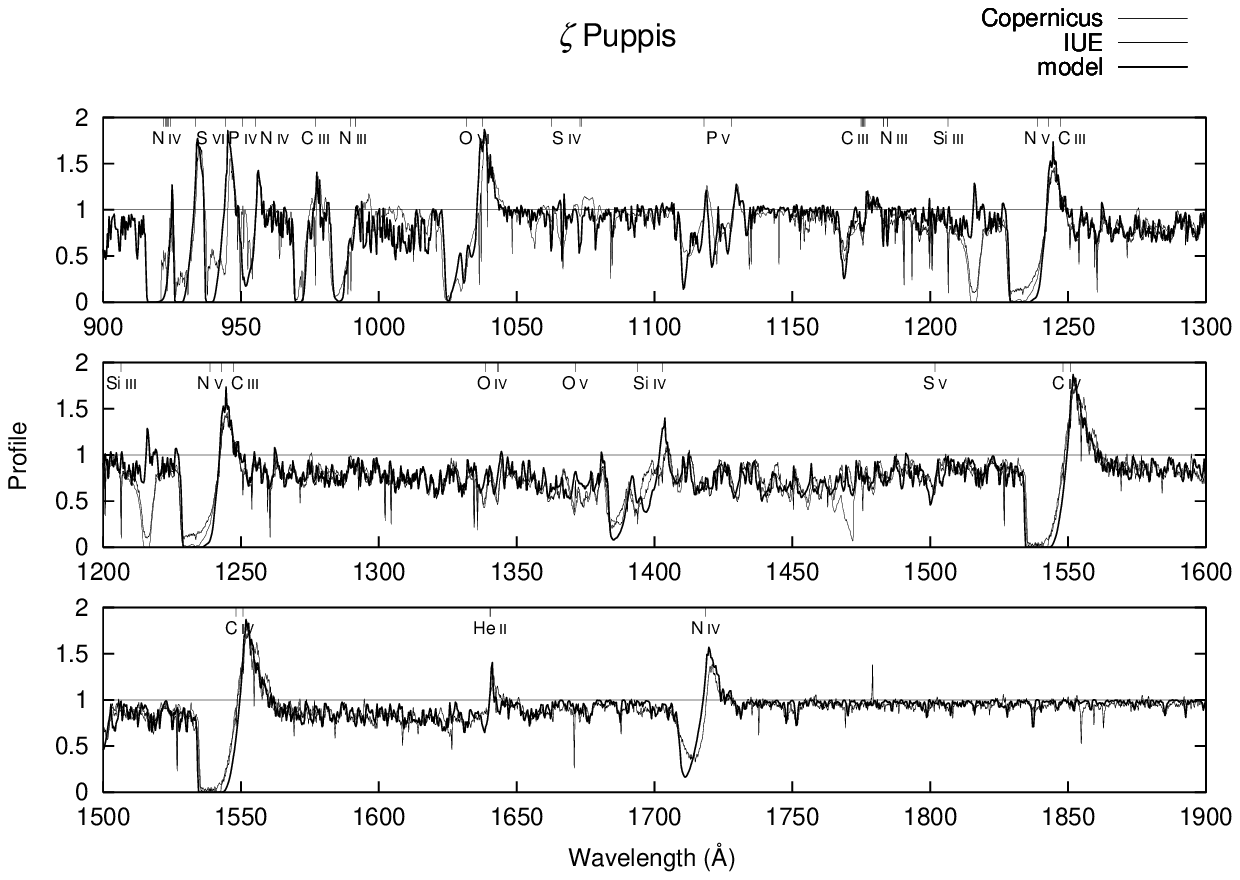}}
\vfill
\myfigure{\includegraphics[width=\textwidth]{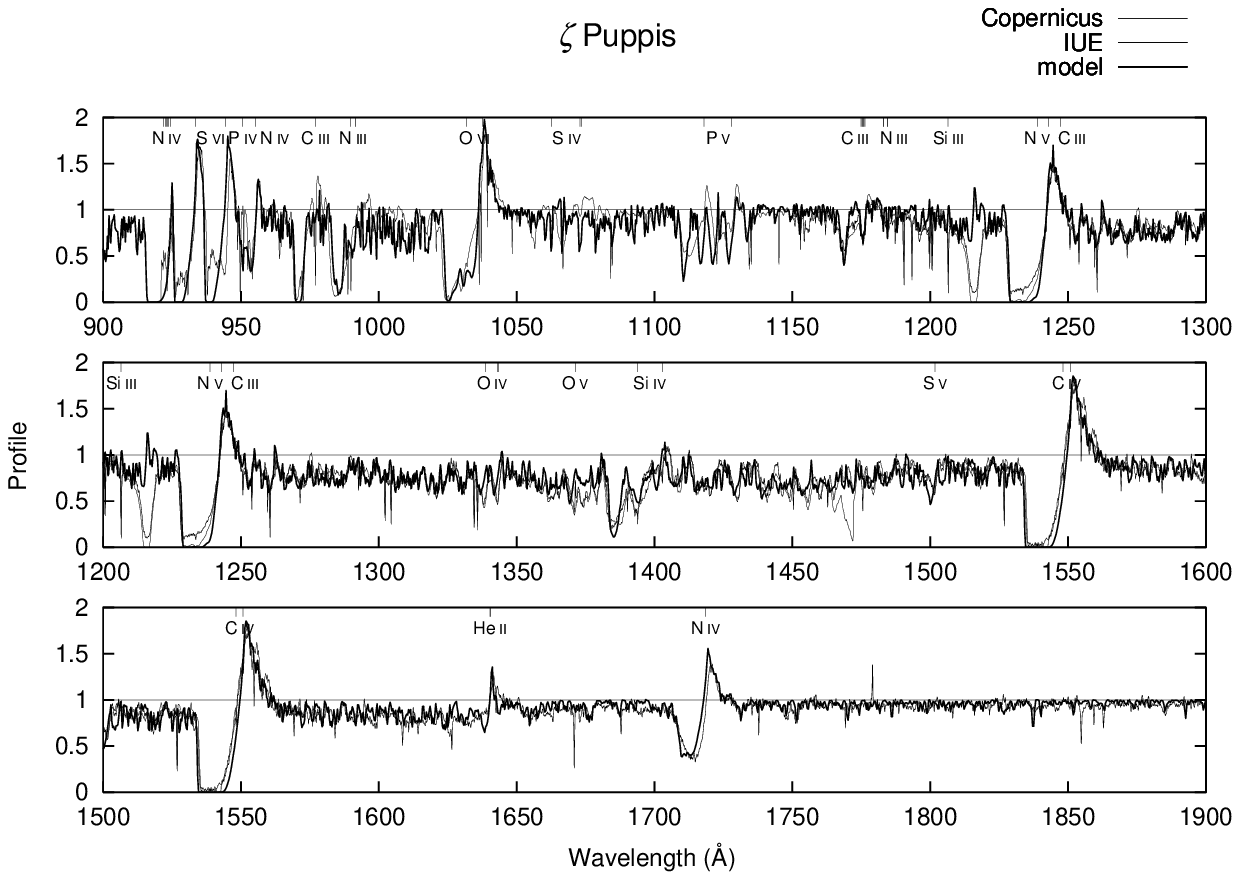}}
\vfill\noindent
Figure 19: {\small Calculated and observed UV spectrum for
$\zeta$~Puppis. The observed spectrum shows the Copernicus and IUE
high-resolution observations, and the calculated state-of-the-art
spectra represent the final models of Pauldrach et\,al.\ 2003. For
a discussion see text.}
% ----------------------------------------------------------------
\pagebreak

% ----------------------------------------------------------------
\myfigure{\includegraphics[width=5.8cm]{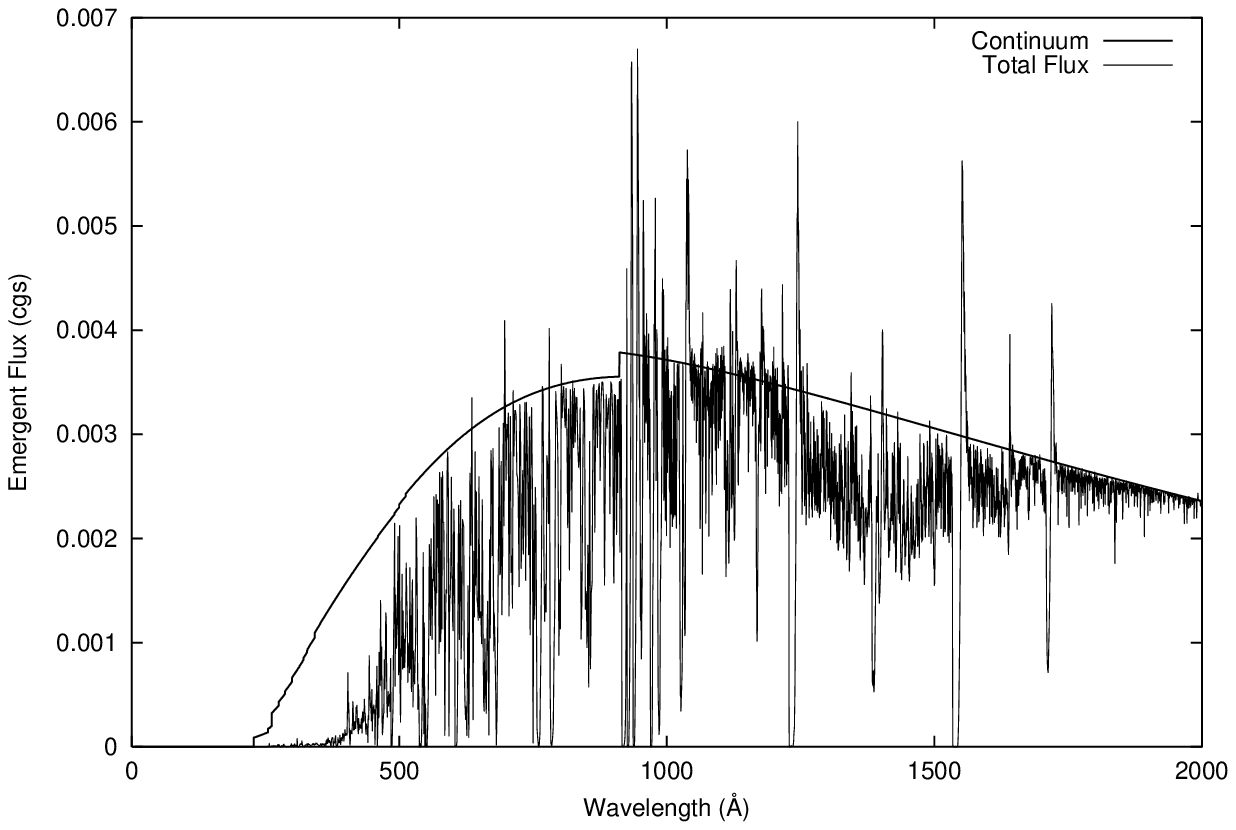}\hfill
\includegraphics[width=5.8cm]{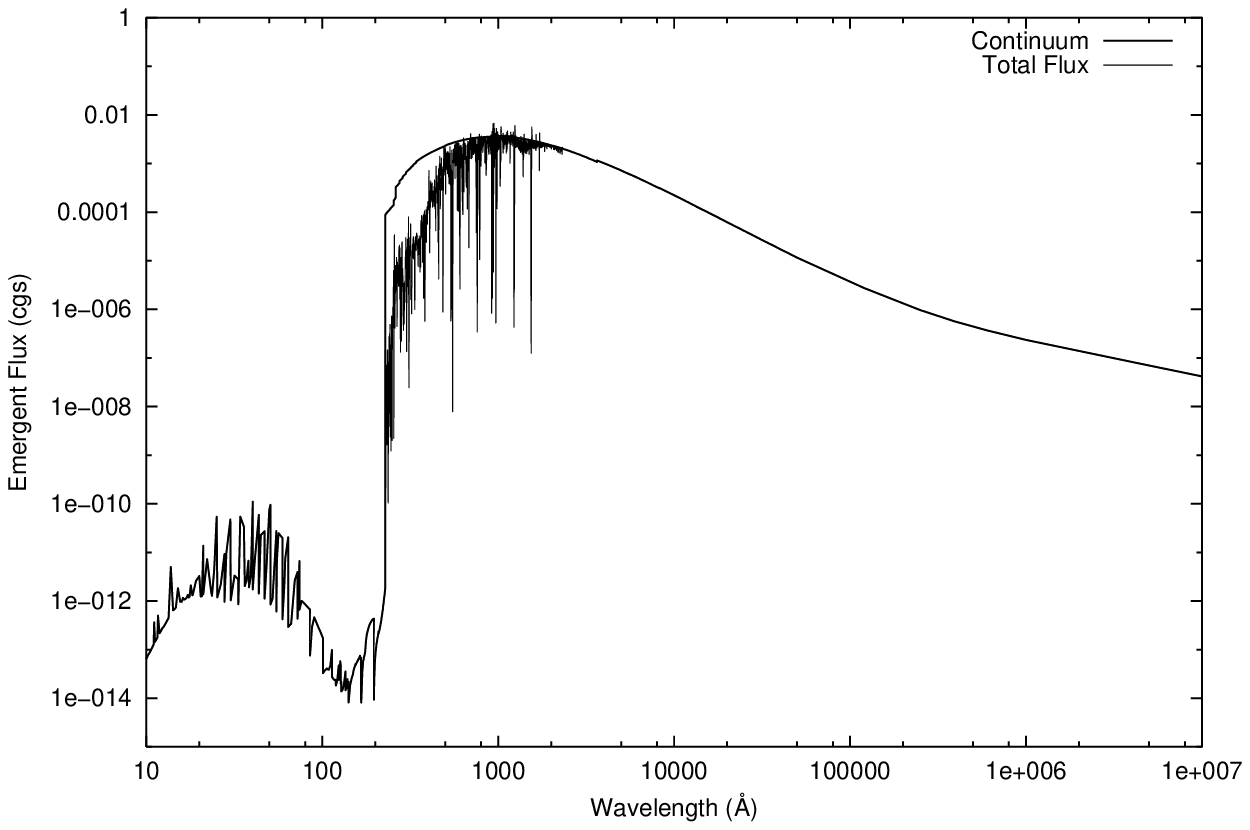}}
Figure 20: {\small Calculated state-of-the-art spectral energy
distribution of the final model of $\zeta$~Puppis of Pauldrach
et\,al.\ 2003. {\em Left\/}:~Linear scale for the flux. {\em
Right\/}:~Logarithmic scale for the flux.}\\
% ----------------------------------------------------------------

\noindent and a simple treatment of emitted radiation from shock
cooling zones. Thus, we conclude that a more consistent treatment
of expanding atmospheres was still needed for a detailed analysis.

This more consistent treatment of expanding atmospheres that has now
been applied and fixes the status quo is based on the improvements
discussed in Section~6.1 (for a more comprehensive discussion see
Pauldrach et\,al.\ 2001 and Pauldrach and Hoffmann~2003). As is
shown in Figure~19, the calculated synthetic spectrum is quite well
in agreement with the spectra observed by IUE and Copernicus. The
small differences observed in the \SiIV\ and the \NIV~lines just
reflect a sensitive dependence on the parameters used to describe
the shock distribution. This is verified by a comparison of the two
panels of Figure~19, where just the shock-distribution has slightly
been changed within the range of uncertainty of the corresponding
parameters. Not only have the stellar and wind parameters of this
object been confirmed by the model on which the synthetic spectrum
is based, but also the abundances of C, N, O, P, Si, S, Fe, and
Ni have been determined. We thus conclude that the present method
of quantitative spectral UV~analysis of Hot Stars leads to models
which can be regarded as being realistic.

Consequently, we consider these kind of quantitative spectral UV
analyses as the ultimate test for the accuracy and the quality of
theoretical ionizing fluxes (cf.~Figure~20), which can thus be used
as spectral energy distributions for the analysis of \HII~regions.

\subsubsection{UV Analysis of the Cool O~Supergiant $\alpha$~Cam}

In Figure~21 it can be seen that the method also works for cool
O~supergiants like $\alpha$~Cam. We recognize again that the
synthetic spectrum is quite well in agreement with the IUE and
Copernicus spectra (for a discussion see Pauldrach et\,al.\ 2001). In
the uppermost panel of the spectrum, simulations of interstellar
lines have additionally been merged with the synthetic spectrum
in order to disentangle wind lines from interstellar lines (for
a discussion of this point see Hoffmann 2002). In this case also,
our new realistic models allowed to determine stellar parameters,
wind parameters, and abundances from the UV~spectra alone. (The
operational procedure of the\linebreak
% ----------------------------------------------------------------
\myfigure{\includegraphics[width=11.4cm]{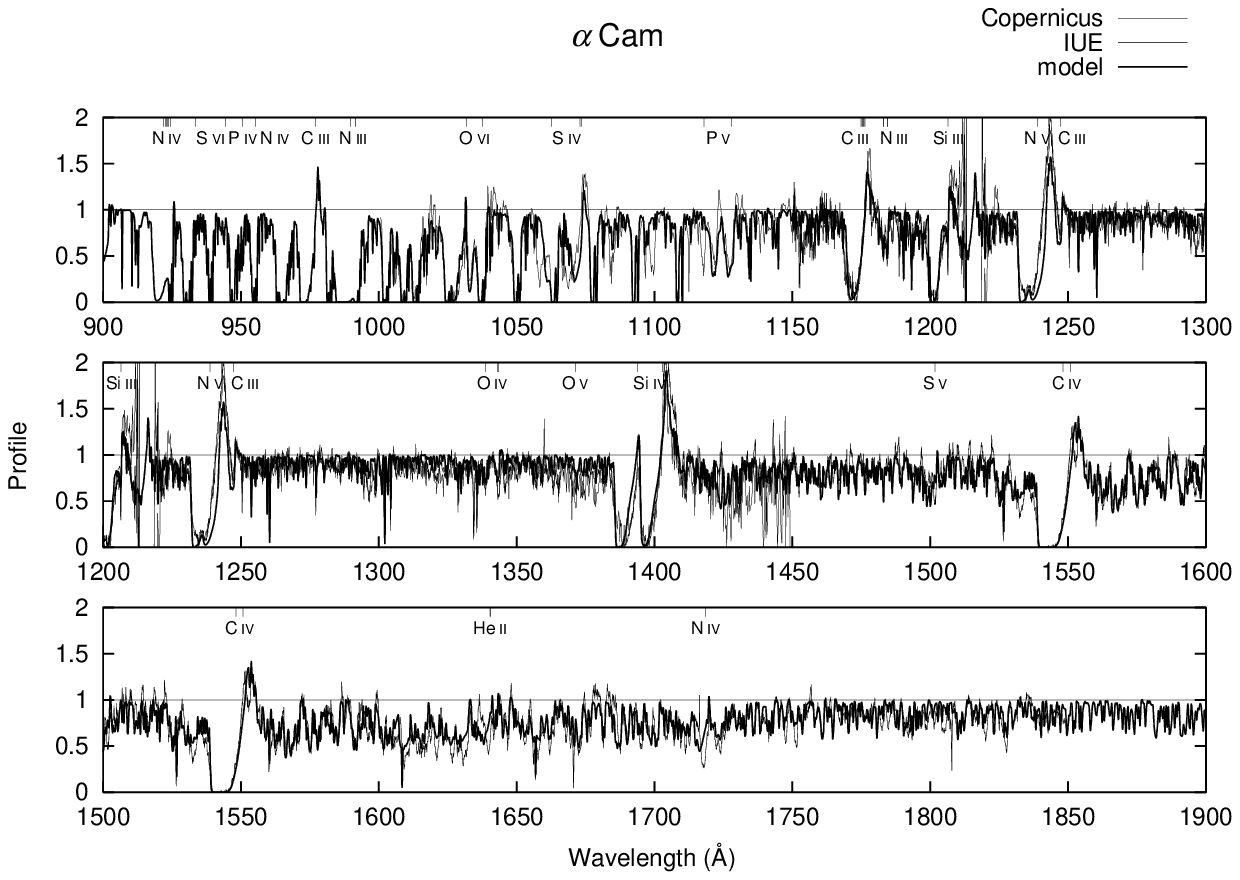}}
Figure 21: {\small Comparison of the best model of Pauldrach
et\,al.\ 2001 with spectra of $\alpha$~Cam observed by IUE and
Copernicus, demonstrating the quality that can be achieved with the
new model generation.}\\
% ----------------------------------------------------------------

\noindent method begins with realistic estimates of $R$, $\Teff$,
$M$, and a set of abundances. With these, the model atmosphere is
solved and the velocity field, the mass loss rate~$\Mdot$, and the
synthetic spectrum is calculated. The parameters are adjusted and
the process is repeated until a good fit to all features in the
observed UV~spectrum is obtained. For a compact description of
this procedure see Pauldrach et\,al.\ 2002). It turned out that
the effective temperature can be determined to within a range of
$\pm 1000$~K and the abundances to at least within a factor of~2.

\subsection{Wind properties of massive O~stars}

As a last point we want to illustrate the significance of the
dynamical parameters of radiation-driven winds. The intrinsic
significance is quite obvious: it is the consistent hydrodynamics
which provides the link between the stellar parameters ($\Teff$,
$M$, $R$) and the wind parameters ($\vinf$, $\Mdot$). Thus, the
appearance of the UV spectrum is determined by the interplay of the
NLTE model and the hydrodynamics. (As was discussed in Section~6.1,
the hydrodynamics is controlled by the line force, which is primarily
determined by the occupation numbers, and the radiative transfer
of the NLTE model, but the hydrodynamics in turn affects the NLTE
model via the density and velocity structure.)

A tool for illustrating the significance of the dynamical parameters
is offered by the so called {\it wind-momentum--luminosity relation}.
This relation is based on two important facts: The first one is
that, due to the driving mechanism of hot stars,\linebreak
% ----------------------------------------------------------------
\myfigure{\includegraphics[width=8.9cm]{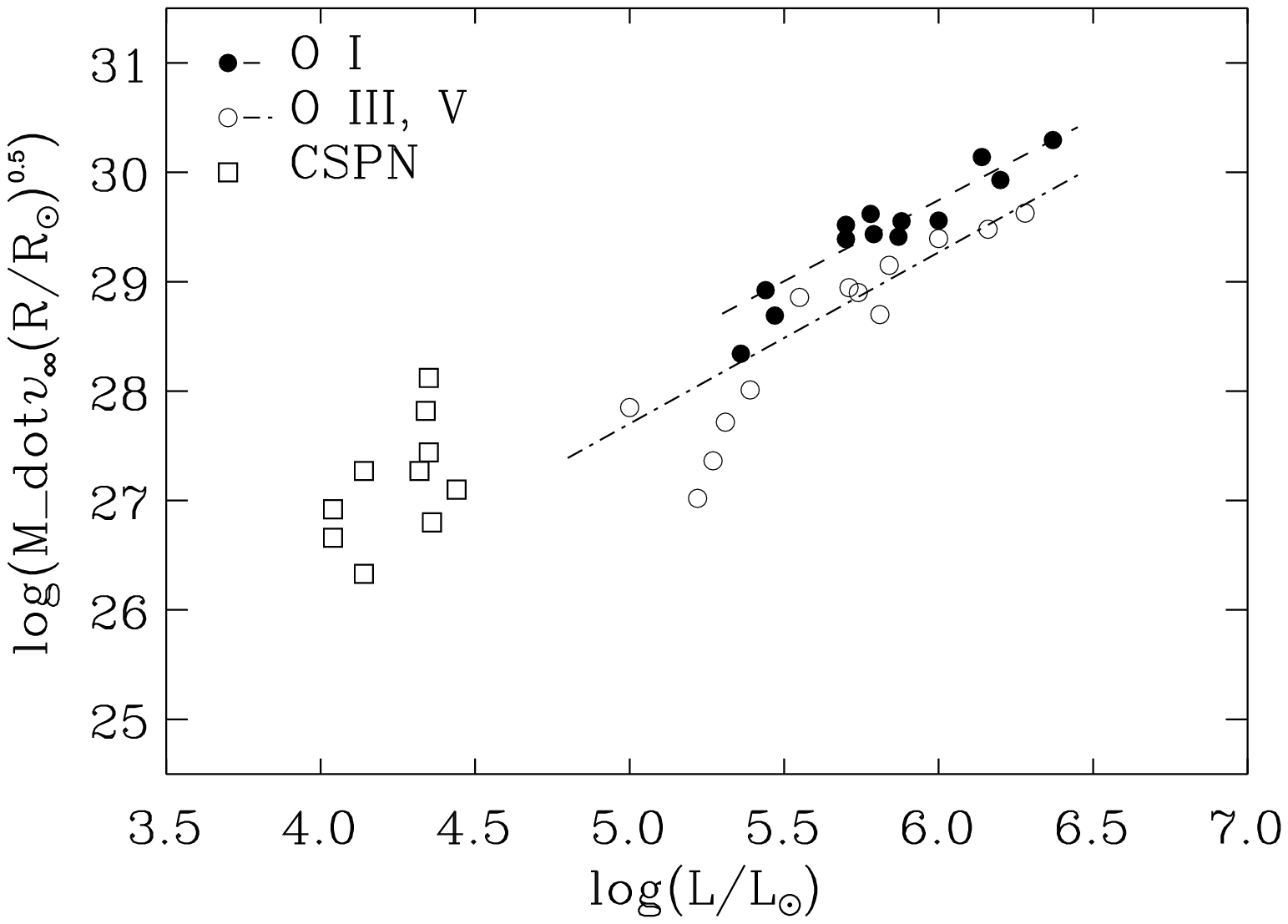}}
Figure 22: {\small The wind-momentum--luminosity relation for
massive O~stars and Central Stars of Planetary Nebulae (CSPNs).
Circles designate the O~star analysis based on H$\alpha$ profiles
by Puls et\,al.\ 1996, and squares that of CSPNs by Kudritzki
et\,al.\ 1997. Figure from Kudritzki and Puls~2000.}\\
% ----------------------------------------------------------------

\noindent the mechanical momentum of the wind flow ($\vinf \Mdot$)
is mostly a function of photon momentum ($L/c$) and is therefore
related to the luminosity.  The second one is that the expression
$\vinf \Mdot R^{1/2} $ is an almost directly observable quantity. As
the shapes of the spectral lines are characterized by the strength
and the velocity of the outflow, the first term of this expression,
the observed terminal velocity, can be measured directly from the
width of the absorption part of the saturated UV resonance lines.
Deducing the mass-loss rate from the line profiles is, however,
more complex, since this requires the calculation of the ionization
balance in advance.  On the other hand, the advantage is that
optical lines like H$\alpha$ can be used for this purpose.  Thus,
the product of the last two terms of the expression $\vinf \Mdot
R^{1/2}$ directly follows from a line fit of H$\alpha$ (cf.~Puls
et\,al.\ 1996). The dynamical parameters obtained in this way are
usually designated as {\it observed wind parameters}. Figure~22
shows that the wind-momentum--luminosity relation indeed exists for
massive O~stars, as they follow the linear relation predicted by
the theory, in a first approximation and for fixed metallicities.
With regard to the spread in wind momenta found by Kudritzki et\,al.\
(1997) for the Central Stars of Planetary Nebulae (lower part of
Figure~22), Pauldrach et\,al.\ (2002, 2003) have given a solution
which, however, is unlikely to be believed yet by the part of the
community working on stellar evolution.

Moreover, it is important to note that as a by-product of this
relation it can be used as an independent tool for measuring
extragalactic distances up to Virgo and Fornax, since it is
independent of the observationally unknown masses (for a recent and
comprehensive review on this subject see Kudritzki and Puls~2000).

Figure~23 shows that the observed behavior of the wind momentum
luminosity relation is represented quite well by our improved
realistic models, particularly in the case of the two supergiants
($\zeta$~Puppis and $\alpha$~Cam) analyzed in Section~7.2.
But,\linebreak
% ----------------------------------------------------------------
\myfigure{\includegraphics[width=11.0cm]{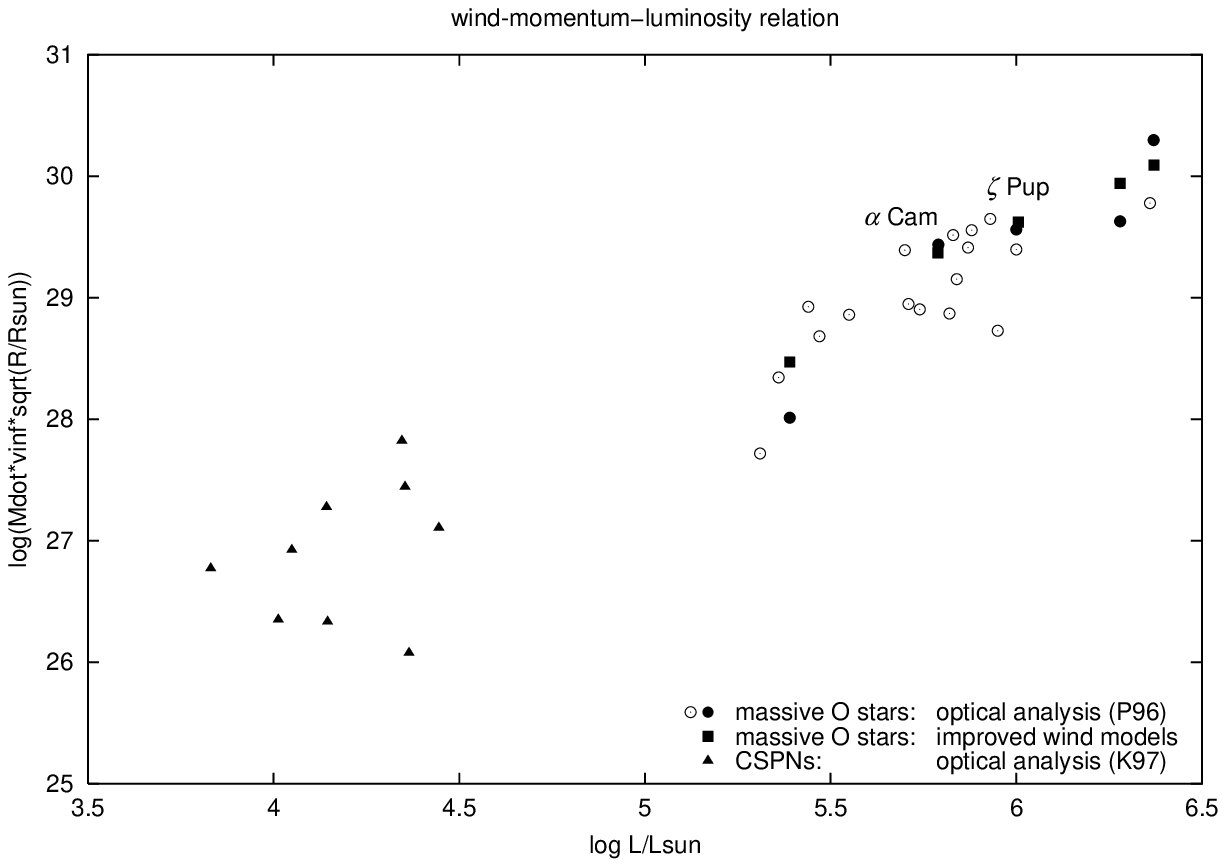}}
Figure 23: {\small The wind-momentum--luminosity relation for
massive O~stars and CSPNs as in Figure~22. Also plotted as filled
squares are the calculated wind momenta for five massive O~stars;
filled circles designate the corresponding observed values
(cf.~Pauldrach et\,al.\ 2002, 2003).}\\
% ----------------------------------------------------------------

\noindent as already pointed out, this relation is independent of
the stellar mass. Thus, in order to verify the statement of the
previous section that the observed stellar and wind parameters are
confirmed by the present models, we also have to investigate the
relations of the individual dynamical parameters.

For this investigation we need to use as input for our models the
same stellar parameters as have been used to obtain the observed
wind parameters.  On the basis of this requirement, Figure~24 (upper
panel) shows that the observed and predicted values of the terminal
velocities ($\vinf$) are in agreement to within 10\,\%. Since $\vinf$
is proportional to the escape velocity ($v_{\rm esc}$)
\[
v_\infty \propto v_{\rm esc} =
\left( \frac{2GM}{R} ( 1 - \Gamma ) \right)^{1/2}
\]
which strongly depends on the mass of the objects, {\it the
mass is determined very accurately by the predicted values\/}
($G$ is the gravitational constant and $\Gamma$ is the ratio of
radiative Thomson acceleration to gravitational acceleration). For
the mass-loss rates we found agreement to within a factor of two,
as shown in Figure~24 (lower panel). Due to the strong relation
between the mass-loss rates and the luminosities ($\Mdot \propto L$),
{\it the luminosities can thus be precisely determined}.

Therefore, computing the wind dynamics consistently with the NLTE
model permits not only the determination of the wind parameters
from given stellar parameters,\linebreak
% ----------------------------------------------------------------
\myfigure{\includegraphics[width=11.2cm]{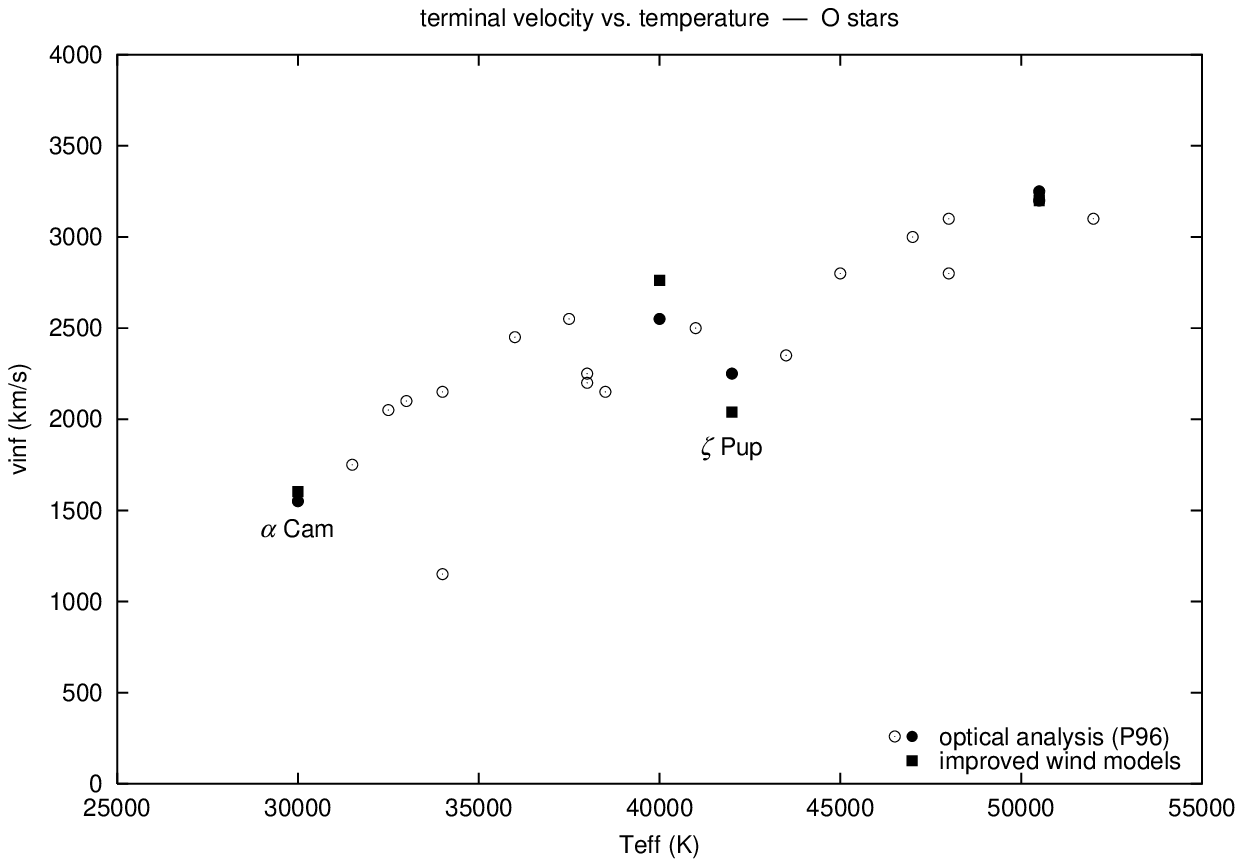}}
\vfill
\myfigure{\includegraphics[width=11.2cm]{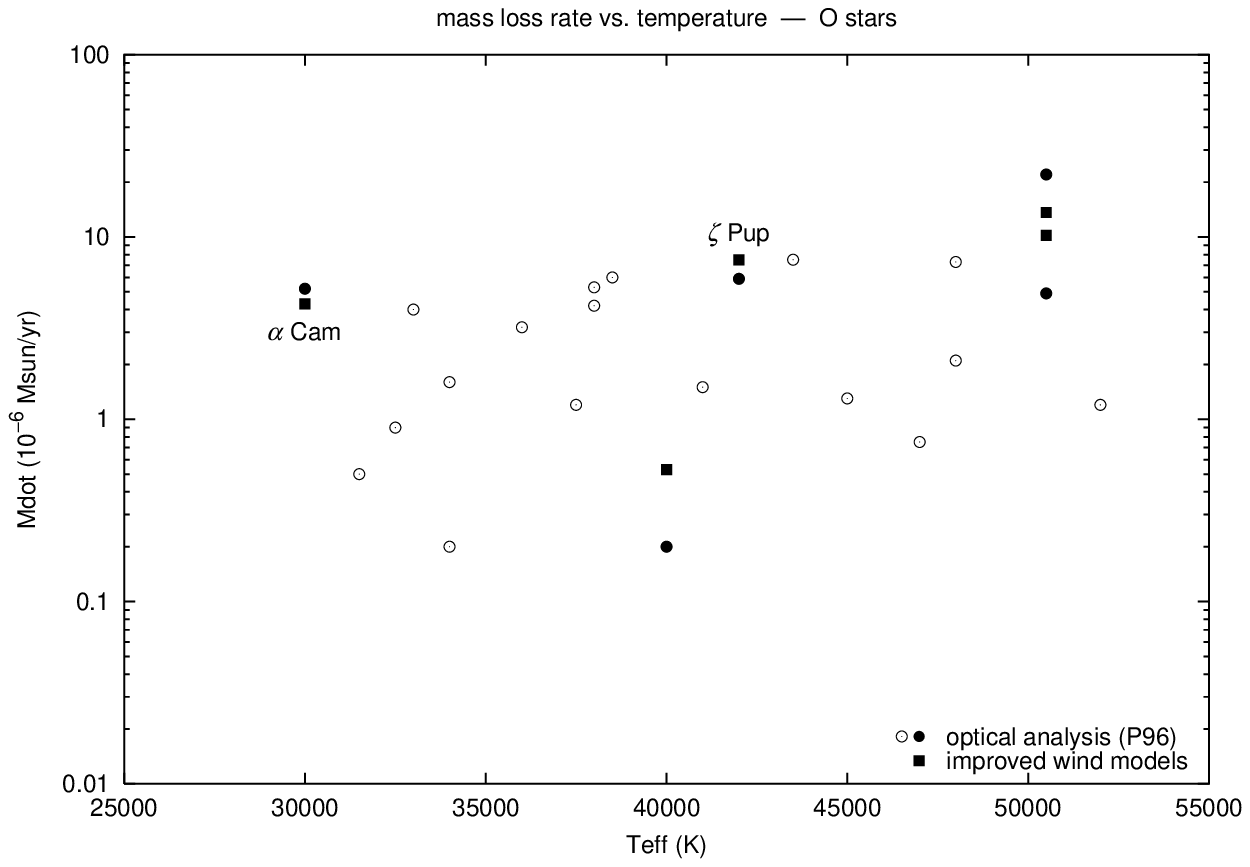}}
\vfill\noindent
Figure 24: {\small {\em Top\/}:~Terminal velocities as a function of
effective temperature for massive O-stars of the same sample
as in Figure~23. {\em Bottom\/}:~Mass-loss
rates as a function of effective temperature for the same
sample. The calculated values for the five massive O~stars are
designated by filled squares; filled circles designate the
corresponding observed values (cf.~Pauldrach and Hoffmann~2003
concerning $\zeta$~Puppis; the other objects have been treated
by Hoffmann and Pauldrach 2002).}
% ----------------------------------------------------------------

\noindent but conversely makes it possible {\it to obtain the stellar
parameters from the observed UV~spectrum alone}. As stated above,
this means that, in principle, {\it the stellar parameters can
immediately be read off by simply comparing an observed UV~spectrum
to a proper synthetic spectrum}.  However, the whole procedure is
not always an easy task, since in most cases a comprehensive grid of
models and some experience will be required to do so. Although the
idea itself is not new (cf.~Pauldrach et\,al.\ 1988 and Kudritzki
et\,al.\ 1992), only the new generation of models has reached the
degree of sophistication that makes such a procedure practicable
instead of purely academic. The corresponding conclusion is that
{\it realistic models are characterized by at least a quantitative
spectral UV~analysis calculated together with a consistent dynamics}.

\smallskip
\section{Conclusions and Outlook}

The \textbf{need} for \textbf{detailed atmospheric models} of
\textbf{Hot Stars} has been motivated in depth and a
\textbf{diagnostic tool} with great astrophysical potential
has been presented.

It has been shown that the models of the {\it new generation\/}
are realistic. They are {\it realistic\/} with regard to a {\it
quantitative spectral UV analysis\/} calculated, for the case
of O~stars, along with consistent dynamics, which, in principle,
allows to read off the stellar parameters by comparing an observed
UV~spectrum to a suitable synthetic spectrum. The new generation
of models has reached a degree of sophistication that makes such
a procedure practicable.

The astronomical perspectives are enormous. Of course, there is still
a large amount of hard and careful astronomical work that needs
to be done. The diagnostic tool for expanding model atmospheres
is in our hands, but this is (as usual) just the starting point;
further elaboration, refinements, and modifications are required
before the results are quantitatively completely reliable.

But this is not the handicap for the future. The present
handicap are future observations. What is most urgently needed is
the \textbf{Next Generation Space Telescope}.

With respect to the topics discussed in this review this telescope
is needed for a variety of reasons:

To tackle the question whether \textbf{Supernovae~Ia} are standard
candles in a cosmological sense realistic models and synthetic
spectra of Type~Ia Supernovae require {\it spectral observations of
objects at mid-redshift\/} and sufficient resolution. The context of
this question is the current surprising result that distant SNe~Ia
at intermediate redshift appear fainter than standard candles in
an empty Friedmann model. Consequently, the current SN-luminosity
distances indicate an accelerated expansion of the universe.

In order to be able to make quantitative predictions
about the influence of very massive, extremely metal-poor
\textbf{Population~III} stars on their galactic and intergalactic
environment one primarily needs observations which can be compared
to the predicted flux spectra that are already available for zero
metallicity and can, in principle, be produced for metallicities
different from zero. Observations with the Next Generation Space
Telescope (NGST) of distant stellar populations at high redshifts
will give us the opportunity to deduce the primordial IMF, thus
allowing a quantitative investigation of the ionization efficiency of
a Top-heavy IMF via {\it realistic spectral energy distributions\/}
of these very massive stars.

NGST observations are also important for determining extragalactic
abundances and population histories of \textbf{starburst galaxies}
from an analysis of \HII~regions. This task also requires energy
distributions of time-evolving stellar clusters, which are calculated
on the basis of a grid of \textit{spectral energy distributions}
of O and early B-type stars. A crucial point with respect to this
is whether the spectral energy distributions of massive stars
are already realistic enough to be used for diagnostic issues
of \HII~regions.  It has been shown in this review that the
\textit{ultimate test} is provided by a comparison of observed
and synthetic UV~spectra of individual massive stars, since the
ionization balance can be traced reliably through the strength and
structure of the wind lines formed throughout the atmosphere.

Furthermore, NGST is important for directly exploiting the diagnostic
perspectives of galaxies with pronounced current star formation. It
has been demonstrated that massive stars dominate the UV wavelength
range in \textbf{star-forming galaxies}, and that therefore the
UV-spectral features of massive O~stars can be used as tracers of
age and chemical composition of starburst galaxies even at high
redshift. This is in particular the case when the flux from these
galaxies is amplified by gravitational lensing through foreground
galaxy clusters; corresponding observations render the possibility
of determining metallicities of starbursting galaxies in the early
universe via {\it realistic UV~spectra} (in the rest frame) of
massive O~stars.

\smallskip
As a final remark it is noted that the solution method of
stationary models for expanding atmospheres is in its present
stage already regarded as a standard procedure towards a realistic
description. Thus, together with an easy-to-use interface and an
installation wizard, the program package \textbf{WM-basic} has
been made available to the community and can be downloaded from
the author's home page.

\subsection*{Acknowledgments}

I wish to thank my colleagues Tadziu Hoffmann and Tamara Repolust
for proofreading the manuscript, and I am grateful to Amiel
Sternberg and Claus Leitherer for providing me with figures
from their publications. This research was supported by the
Sonderforschungsbereich~375 of the Deutsche Forschungsgemeinschaft,
and by the German-Israeli Foundation under grant I-551-186.07/97.

\subsection*{References}

{\small

\bref
Baldwin, J.A., Ferland, G.J., Martin, P.G. et\,al. 1991, ApJ 374, 580

\bref
Bromm, V., Coppi, P.S., Larson, R.B. 1999, ApJ 527, L5

\bref
Bromm, V., Kudritzki, R.P., Loeb, A. 2001, ApJ 552, 464

\bref
Carr, B.J., Bond, J.R., Arnett, W.D. 1984, ApJ 277, 445

\bref
Cassinelli, J., Olson, G. 1979, ApJ 229, 304

\bref
Castor, J.I., Abbott, D.C., Klein, R. 1975, ApJ 195, 157

\bref
Conti, P.S., Leitherer, C., Vacca, W.D. 1996, ApJ 461, L87

\bref
El Eid, M.F., Fricke, K.J., Ober, W.W. 1983, A\&A 119, 54

\bref
Fan, X. et\,al. 2000, AJ 120, 1167

\bref
Feldmeier,~A., Kudritzki,~R.-P., Palsa,~R., Pauldrach,~A.\,W.\,A., Puls,~J. 1997, A\&A 320, 899

\bref
Genzel, R. et\,al. 1998, ApJ 498, 579

\bref
Giveon, U., Sternberg, A., Lutz, D., Feuchtgruber, H., Pauldrach,~A.\,W.\,A. 2002, ApJ 566, 880

\bref
Gunn, J.E., Peterson, B.A. 1965, ApJ 142, 1633

\bref
Hamann, W.R. 1980, A\&A 84,342

\bref
Harnden, F.R., Branduardi, G., Elvis, M. et\,al. 1979, ApJ 234, L51

\bref
Hillebrandt, W., Niemeyer, J.C. 2000, ARA\&A 38, 191

\bref
Hoffmann,~T.L. 2002, Ph.D. thesis, LMU Munich

\bref
Hoffmann,~T.L., Pauldrach,~A.W.A. 2002, IAU Symposium 209, in press

\bref
Hummer, D.G., Rybicki, G.B. 1985, ApJ 293, 258

\bref
Kirshner, R.P., Jeffery, D.J., Leibundgut, B., et\,al. 1993, ApJ 415, 589

\bref
Kudritzki, R.P., Hummer, D.G., Pauldrach, A.W.A., et\,al. 1992, A\&A 257, 655

\bref
Kudritzki, R.P., M\'endez, R.\,H., Puls, J., McCarthy, J.\,K. 1997, in IAU Symp.\,180, Planetary Nebulae, eds.\ H.J.\ Habing \& H.J.G.L.M.\ Lamers, p.\ 64

\bref
Kudritzki, R.P., Puls, J. 2000, ARA\&A 38, 613

\bref
Kudritzki, R.P. 2002, ApJ in press

\bref
Kunth, D., Mas-Hesse, J.M., Terlevich, R., et\,al. 1998, A\&A 334, 11

\bref
Kurucz, R.L. 1992, Rev.\ Mex.\ Astron.\ Astrof.\ 23, 181

\bref
Lamers, H.J.G.L.M., Morton, D.C. 1976, ApJ Suppl.\ 32, 715

\bref
Larson, R.B. 1998, MNRAS 301, 569

\bref
Leibundgut, B. 2001, ARA\&A 39, 67

\bref
Leitherer, C., Le{\~a}o, J., Heckman, T.M., Lennon, D.J., Pettini, M., Robert, C. 2001, ApJ 550, 724

\bref
Lucy, L.B., Solomon, P. 1970, ApJ 159, 879

\bref
Lucy, L.B., White, R. 1980, ApJ 241, 300

\bref
Lutz, D. et\,al. 1996, A\&A 315, 137

\bref
Loeb, A. 1998, in ASP Conf.\ Ser.\ 133, 73

\bref
Milne, E.A. 1926, MNRAS 86, 459

\bref
Morton, D.C., Underhill, A.B. 1977, ApJ Suppl.\ 33, 83

\bref
Nomoto, K., Thielemann, F.-K., Yokoi, K. 1984, ApJ 286, 644

\bref
Nugent, P., Baron, E., Branch, D., et\,al. 1997, ApJ 485, 812

\bref
Oey, M.S., Massey, P. 1995, ApJ 452, 210

\bref
Owocki, S., Castor, J., Rybicki, G. 1988, ApJ 335, 914

\bref
Pauldrach,~A.W.A., Puls, J., Kudritzki, R.P. 1986, A\&A 164, 86

\bref
Pauldrach,~A.W.A. 1987, A\&A 183, 295

\bref
Pauldrach,~A.W.A., Herrero,~A. 1988, A\&A 199, 262

\bref
Pauldrach, A.W.A., Puls, J., Kudritzki R.-P., et\,al. 1988, A\&A 207, 123

\bref
Pauldrach,~A.W.A., Feldmeier,~A., Puls,~J., Kudritzki,~R.-P. 1994b, in Space Sci. Rev. 66, 105

\bref
Pauldrach,~A.W.A., Kudritzki,~R.-P., Puls,~J., et\,al. 1994, A\&A 283, 525

\bref
Pauldrach,~A.W.A., Duschinger,~M., Mazzali,~P.A., et\,al. 1996, A\&A 312, 525

\bref
Pauldrach,~A.W.A., Lennon,~M., Hoffmann,~T.L., et\,al. 1998, in: Proc.\ 2nd Boulder-Munich Workshop, PASPC 131, 258

\bref
Pauldrach,~A.W.A., Hoffmann, T.L., Lennon, M. 2001, A\&A 375, 161

\bref
Pauldrach,~A.W.A., Hoffmann,~T.L., Mendez, R.H. 2002, IAU Symposium 209, in press

\bref
Pauldrach,~A.W.A., Hoffmann,~T.L., Mendez, R.H. 2003, A\&A, in press

\bref
Pauldrach,~A.W.A., Hoffmann,~T.L. 2003, A\&A, in press

\bref
Perlmutter, S., Aldering, G., Goldhaber, G., et\,al. 1999, ApJ 517, 565

\bref
Pettini, M., Kellogg, M., Steidel, C.C., et\,al. 1998, ApJ 508, 539

\bref
Pettini, M., Steidel, C.C., Adelberger, K.L., Dickinson, M., Giavalisco, M. 2000, ApJ 528, 96

\bref
Puls, J. 1987, A\&A 184, 227

\bref
Puls, J., Pauldrach,~A.W.A. 1990, PASPC 7, 203

\bref
Puls, J., Kudritzki R.-P., Herrero A., Pauldrach,~A.W.A., Haser, S.M., et\,al. 1996, A\&A 305, 171

\bref
Reinecke, M., Hillebrandt, W., Niemeyer, J.C. 1999, A\&A 374, 739

\bref
Riess, A.G., Filippenko, A.V., Challis, P., et\,al. 1998. Astron. J. 116, 1009

\bref
Riess, A.G., Kirshner, R.P., Schmidt, B.P., et\,al. 1999, Astron. J. 117, 707

\bref
Rybicki, G.B. 1971, JQSRT 11, 589

\bref
Rubin, R.H., Simpson, J.P., Haas, M.R., et\,al. 1991, PASP 103, 834

\bref
Saha, A., Sandage, A., Thim, F., Labhardt, L., Tammann, G.A., Christensen, L., Panagia, N., Macchetto, F. 2001, ApJ, 551, 973

\bref
Sauer, D., Pauldrach,~A.W.A. 2002, Nucl. Astrophys., MPA/P13

\bref
Seward, F.D., Forman, W.R., Giacconi, R., et\,al. 1979, ApJ 234, L55

\bref
Schaerer,~D., de Koter,~A. 1997, A\&A 322, 598

\bref
Sellmaier, F.H., Yamamoto, T., Pauldrach,~A.W.A., Rubin, R.H. 1996, A\&A, 305, L37

\bref
Simpson, J.P., Colgan, S.W.J., Rubin, R.H., et\,al. 1995, ApJ 444, 721

\bref
Sobolev, V. 1957, Sov. A\&A J. 1, 678

\bref
Steidel, C.C., Giavalisco, M., Pettini, M., Dickinson, M., Adelberger, K.L. 1996, ApJ 462, L17

\bref
Sternberg, A., Hoffmann, T.L., Pauldrach~A.W.A. 2002, ApJ, in press

\bref
Thornley, M.D., F\"{o}rster-Schreiber, N.M., Lutz, D., et\,al. 2000, ApJ 539, 641

\bref
Walborn, N.R., Nichols-Bohlin, J., Panek, R.J. 1985, IUE Atlas of O-Type Spectra from 1200 to 1900{\AA} (NASA RP-1155)

}

\vfill
\pagebreak

\hspace*{1cm}
\thispagestyle{empty}
\vfill
\pagebreak

\end{document}